\documentclass[
  aps,
  prb,
  reprint,
  superscriptaddress,
  amsmath,amssymb
]{revtex4-2}

\usepackage{graphicx}
\usepackage{dcolumn}
\usepackage{bm}
\usepackage{amssymb}
\usepackage{amsfonts}

\usepackage[colorlinks=true, linkcolor=blue, citecolor=blue, urlcolor=blue]{hyperref}
\usepackage{cleveref}

\usepackage{amsmath,mathtools}
\DeclareMathOperator{\atantwo}{atan2}

\usepackage[T1]{fontenc}
\usepackage{url}
\usepackage[margin=25mm]{geometry}

\newgeometry{top=2cm, left=1.65cm,right=1.65cm}

\begin{document}

\title{Controllable non-reciprocity in multi-sphere loaded chiral resonator}

\author{Maxime Ardisson}
\email{maxime.ardisson@imt-atlantique.fr}
\affiliation{IMT Atlantique, Technopole Brest-Iroise, CS 83818, 29238 Brest Cedex 3, France}
\affiliation{Lab-STICC (UMR 6285), CNRS, Technopole Brest-Iroise, CS 83818, 29238 Brest Cedex 3, France}

\author{Guillaume Bourcin}
\affiliation{KWAN-TEK, 1 rue Galilée, Espace Innova, 56270 Ploemeur, France}

\author{Julien Haumant}
\affiliation{Elliptika, 29804 Brest Cedex 9, France}

\author{Romain Lebrun}
\author{Isabella Boventer}
\affiliation{Laboratoire Albert Fert, Thales, Université Paris-Saclay, Palaiseau 91767, France}

\author{Vincent Castel}
\affiliation{IMT Atlantique, Technopole Brest-Iroise, CS 83818, 29238 Brest Cedex 3, France}
\affiliation{Lab-STICC (UMR 6285), CNRS, Technopole Brest-Iroise, CS 83818, 29238 Brest Cedex 3, France}

\begin{abstract}
Cavity magnonics explores the hybridization of photons and magnons within microwave resonators. One of the hallmarks of these systems is their ability to exhibit non-reciprocity, which is a key feature for radio frequency (RF) applications. One way to control non-reciprocal behaviors in cavity magnonics is the design of chiral cavities that allow selective coupling between photons and magnons depending on their polarization. However, a built-in chiral platform to harness and control non-reciprocity remains to be achieved. Here, we experimentally demonstrate controllable non-reciprocity (with an absolute isolation ratio reaching 46 dB) in a chiral resonator loaded with multiple yttrium iron garnet spheres. We develop a theoretical model of the S-parameters based on input-output formalism which highlights the links between the phases occurring in the system and its non-reciprocal behavior. Controllable non-reciprocity in cavity magnonics could enable the development of programmable isolators, circulators, and RF switches with improved performance. Such developments could pave the way toward more versatile and scalable information processing systems.
\end{abstract}

\maketitle

\section{\label{sec:level1} Introduction}

More than a decade after the initial experimental observations of cavity-magnon-polaritons (CMPs) \cite{Huebl2013}, the field of cavity magnonics has matured as a rich hybrid research field. CMPs are hybrid quasiparticles arising from the strong coupling of spin waves and photons in a microwave cavity, where spin waves are defined as dynamic magnetization excitations in a collective spin ensemble \cite{Tabuchi2014, Huebl2013, Zhang2014}. In the coherent regime, CMPs can be characterized by a level repulsion in the frequency spectrum between magnon (quanta of spin waves) and photon modes called anticrossing \cite{Zhang2014}. Motivated by the ability of high spin density materials to strongly couple with light and allow coherent information exchange at room temperature \cite{Soykal2010,Huebl2013,Tabuchi2014,Zhang2014}, cavity magnonics systems have been extended to other coupling phenomena e.g. ultrastrong coupling \cite{Flower2019,Bourcin2023}, dissipative coupling \cite{Harder2018,Zhang2017,Rao2021} or cavity-mediated spin-spin coupling \cite{Lambert2016,ZareRameshti2018}. In addition, several cavity magnonics systems have been shown to exhibit non-reciprocal behavior \cite{Wang2019,Zhang2020,Zhong2022}, which is a property of significant interest for classical communication devices such as isolators or circulators, paving the way for magnonic coherent information processing. Chirality on the other hand plays a crucial role in governing wave propagation in a wide variety of physical systems such as photonics \cite{Lodahl2017, Coles2016}, plasmonics \cite{Zhang2011} or condensed matter physics \cite{Zaz2025,Litvinov2024}. It is important to note that here, the concept of chirality goes beyond its traditional geometrical interpretation to encompass the helicity and handiness of the spin. In cavity magnonics, chirality arises from the interplay between the orientation of the spins in magnetically ordered materials and the spin angular momentum associated to the cavity modes \cite{Yu2020, Bliokh2017}. Chiral interactions find applications in optomagnonics \cite{Hou2025}, dissipative cavity magnonic systems \cite{Zhao2025} and topology \cite{https://doi.org/10.48550/arxiv.2412.10888}. However, previous systems exploiting chirality for non-reciprocity in cavity magnonics were ultimately limited by planar integration and cavity volume, achieving isolation ratio of $20$ dB \cite{Bourhill2023}, here we report an isolation ratio of $50$ dB. In this study, we address the non-reciprocity of a system constituted of ferrimagnetic insulators coupled to chiral modes inside a toroidal-shaped substrate integrated waveguide (SIW) \cite{Yu2020, Bourhill2023}. According to Maxwell's equations, at certain special positions inside the resonator, the polarization of the photon modes is linked to their propagation direction. Chiral interaction is achievable when ferrimagnetic insulators, such as yttrium iron garnet (YIG) spheres, are placed on these special positions and couple to the cavity modes by following the resonant coupling condition: $\omega_{cavity}\equiv \omega_{magnon}$. If these conditions are fulfilled, the magnon mode will resonantly couple to photons with the same polarization.

While some studies explored specifically the theoretical \cite{Yu2020,TaoYu2020, Yang:24} and experimental \cite{Bourhill2023} aspects of chiral microwave-magnon coupling, a global understanding of the non-reciprocity through its dependence on the position and number of YIG spheres in an integrable and functional system is still lacking. In this work, we present an integrated platform for non-reciprocity control and the study of chiral interactions in the context of cavity magnonics.

In Sec. \hyperref[Sec2]{II}, we first introduce the design of our cavity and review its dynamics for both cases: empty and loaded with YIG spheres. In Sec. \hyperref[Sec3]{III}, we address the theoretical model of our system detailing the calculation of the transmission parameter from the system's Hamiltonian. We then discuss the observation of strong coupling according to the number of YIG spheres inside the cavity in Sec. \hyperref[Sec4]{IV} before presenting our results on non-reciprocity in Sec. \hyperref[Sec5]{V}.

\section{Chiral resonator}\phantomsection\label{Sec2}

\begin{figure}
\includegraphics[width=\linewidth]{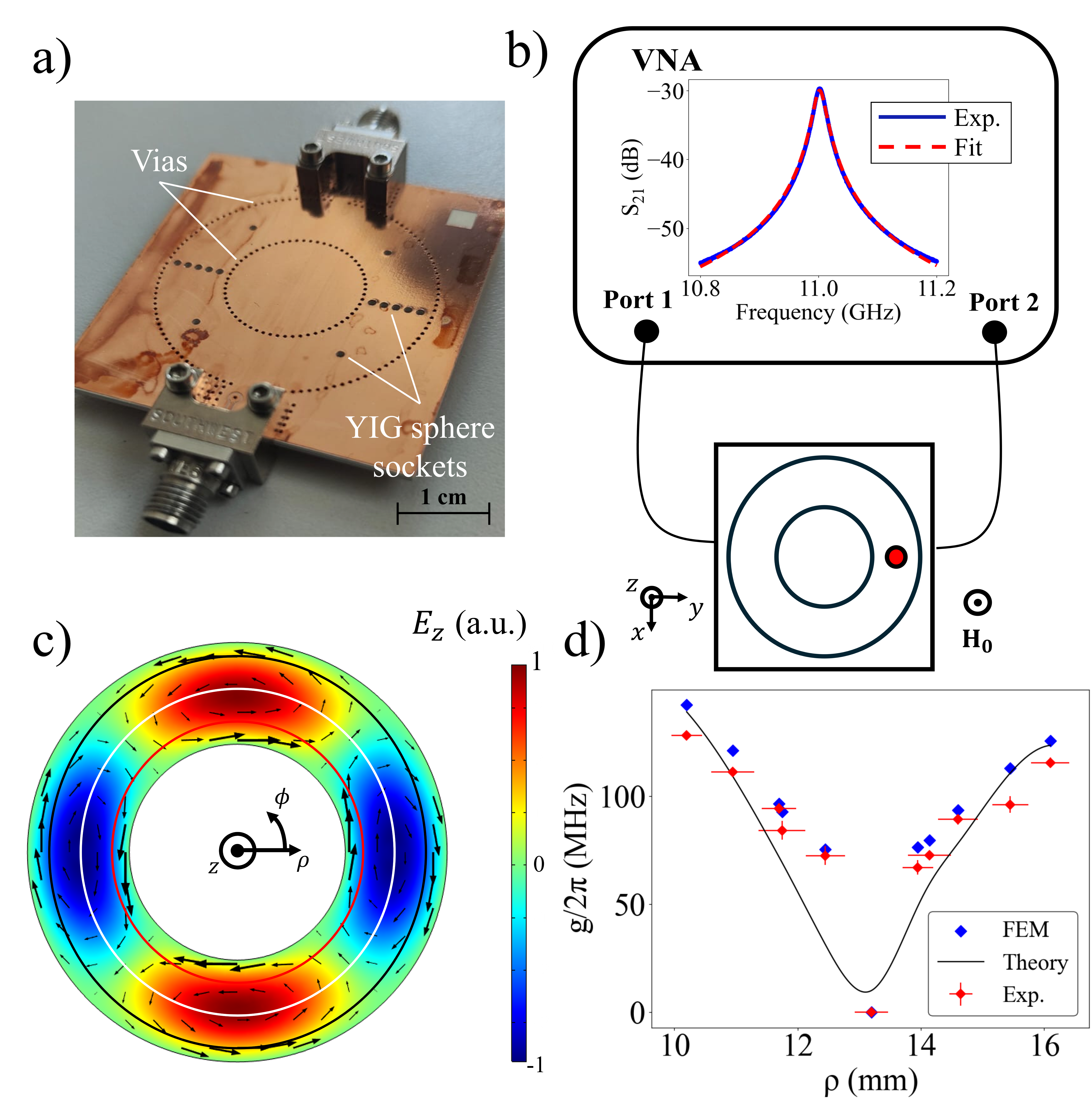}
\caption{SIW chiral resonator for magnon-photon coupling. \textbf{a)} Picture of the SIW chiral cavity. \textbf{b)} Sketch of the measurement system along with the orientation of the DC magnetic field $\text{H}_0$. The inset shows the transmission amplitude ($S_{21}$ in dB) of the empty resonator in blue, with resonant frequency: $\frac{\omega\textsubscript{c}}{2 \pi}= 11$ GHz, intrinsic damping rate: $\frac{\kappa\textsubscript{c}}{2 \pi}=19.7$ MHz and loaded \textit{Q}-factor: $Q_L=560\pm35$ extracted from the Lorentzian fit curve in red. \textbf{c)} Simulated distribution of $E_z$ at $11$ GHz with the orientation of the cavity RF magnetic field indicated in black arrows. The three circles denote the special positions of the resonator: in red $\rho\textsubscript{-}$, white $\rho\textsubscript{0}$, and black $\rho\textsubscript{+}$. \textbf{d)} Radial dependency of the coupling strength for $N=1$ on the azimuthal position $\phi=0$ measured experimentally (Exp. in red), computed by finite element method (FEM in blue) and $|g_+-g_-|$ (Theory in black). The error bars take into account the uncertainty on the YIG sphere position.}
\label{FIG_1}
\end{figure}

The design of our microwave cavity is based on SIW technology which consists in a dielectric (Rogers RO3003\textsuperscript{®} with dielectric constant: $\epsilon\textsubscript{r}=3$) wedged between two metallic layers, it combines both usual waveguide structures with planar integration technology. An array of vias with 250 $\mu m$ radius encloses the electromagnetic field \cite{Zhang2020} inside the substrate and forms the toroidal shape of our resonator with height, inner and outer radii: $h$ = 1.52 mm, $R\textsubscript{1}$ = 9 mm and $R\textsubscript{2}$ = 17.5 mm respectively (see Figure \ref{FIG_1} \textbf{a)}). Appendix \hyperref[appendixA1]{A1} provides additional details on the device fabrication. The magnetic field inside the resonator can be written in the polar basis as $\textbf{H}(\rho,\phi)=H_{\rho}\textbf{e}_{\rho}+H_{\phi}\textbf{e}_{\phi}$, with $(\rho, \phi)$ the polar coordinates, $H\textsubscript{$\rho$}$ and $H\textsubscript{$\phi$}$ the radial and azimuthal components of the cavity magnetic field respectively, detailed in Appendix \hyperref[appendixA2]{A2}. The microwave resonator is holding a frequency degenerate eigenmode supporting both clockwise (CW) and counter-clockwise (CCW) propagation direction at resonant frequency $\frac{\omega\textsubscript{c}}{2\pi}=11$ GHz (see Figure \ref{FIG_1} \textbf{b)}).

Several holes with a diameter of 1.1 mm and a depth of 1.3 mm were drilled at specific $\rho$ positions to host the YIG spheres as shown in Figure \ref{FIG_1} \textbf{a)}. The resonator is loaded with $N=\{1,2,3,4\}$ YIG spheres of $1$ mm diameter \cite{ferrisphere}, the center of each sphere is placed at $z=h/2$ in the $\{\rho, \phi\}$ plane of the cavity. Moreover, the use of the SIW architecture, based on dielectric materials with a relative permittivity higher than that of air ($\epsilon_r$ = 3 in our case), allows a significant reduction in the cavity’s physical dimensions for a given resonant frequency \cite{Rao2021,Bourhill2023}. As a result, integrating a 1 mm YIG sphere into this more compact dielectric-filled cavity leads to a higher magnon-photon coupling strength compared to air-filled cavities operating at the same frequencies. This approach yields a 56\% enhancement in coupling strength at constant YIG volume, compared to previous studies based on circular air-filled cavities \cite{Yu2020, Bourhill2023}. In this work, we will focus on the uniform precession mode of magnetization known as the Kittel mode \cite{Kittel1958}. At specific azimuthal lines, the magnetic field of the transverse electric (TE) modes is circularly polarized, either right-hand (position $\rho=\rho_+$ with $m>1$) or left-hand polarized (position $\rho=\rho_-$ with $m<1$). We define $\rho_0$ as the special position for which the linearly polarized $E_z$ field is maximum. The three special positions $\rho_0$ and $\rho_{\pm}$ are displayed in Figure \ref{FIG_1} \textbf{c)}. A sphere with magnetization set along the $z$-axis by the DC magnetic field $\text{H}_0$ located at one of the chiral lines ($\rho_-$ or $\rho_+$) in the resonator will couple only to the mode with the same polarization \cite{Nambu2020} and break the degeneracy between the two cavity modes. The magnon-photon coupling is totally chiral on position $(\rho=\rho_{\pm}, \phi=0)$, not chiral on $(\rho=\rho_{0}, \phi)$ and partially chiral everywhere else \cite{TaoYu2020, Yu2020}.

\begin{figure*}
    \centering
    \includegraphics[width=\linewidth]{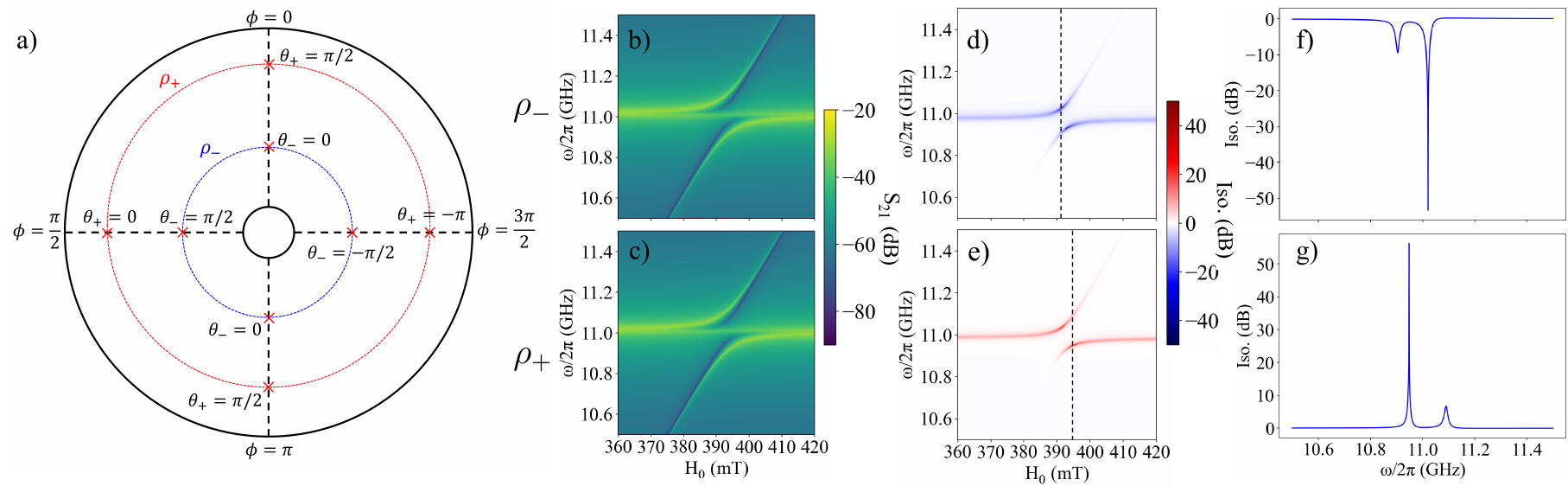}
    \caption{Theoretical model of non-reciprocal chiral coupling for a single YIG sphere. \textbf{a)} Schema of the coupling phases in the polar plane of the chiral cavity with $\phi$ the azimuthal coordinate and $\theta_s$ the coupling phase of the system. The blue and red dashed lines correspond to the chiral lines $\rho_-$ and $\rho_+$ respectively. \textbf{b-c)} Transmission spectrum $S_{21}$ as a function of the applied field \textbf{H} obtained by using equation (\ref{S21}) for a YIG sphere located at $(\rho_-,0)$ \textbf{(b)} and $(\rho_+,0$) \textbf{(c)}. \textbf{d-e)} Isolation ratio spectrum ($S_{21}-S_{12}$ in dB) as a function of $\text{H}_0$ obtained by using Appendix \hyperref[appendixA2]{A2}. Exemplary spectrum at a fixed magnetic field (corresponding to the black dashed lines of \textbf{d-e)}) are shown in \textbf{f)} and \textbf{g)}.}
    \label{FIG_2}
\end{figure*}

For a sphere, the field dependence of the Kittel mode resonance frequency is defined as $\omega_K=\gamma \text{H}_0$, where $\gamma=28$ GHz/T is the gyromagnetic ratio and $\text{H}_0$ the externally applied static magnetic field. In order to experimentally observe CMPs in our system, we conduct spectroscopy measurements using a 2-port Vector Network Analyzer (VNA) while applying $\text{H}_0$ to tune the Kittel resonance frequency of the spheres as sketched in Figure \ref{FIG_1} \textbf{b)} and achieve the resonant coupling condition $\omega_m \equiv \omega_K$. The VNA is connected to the resonator via coaxial cables and measures the $2\times2$ \textit{S}-matrix components.

\section{Theoretical model}\phantomsection\label{Sec3}

The Hamiltonian of the system can be written as the sum of free harmonic oscillators $\hat{H}_{free}$ and an interaction term $\hat{H}_{int}$. The free Hamiltonian reads:

\begin{equation}
    \hat{H}_{free}=\sum_m\hbar\Tilde{\omega}_m\hat{\alpha}_{m}^\dag\hat{\alpha}_{m} + \sum^N_{l=1}\hbar\Tilde{\omega}_K\hat{m}_{l}^\dag\hat{m}_{l},
    \label{Hfree}
\end{equation}

the first term corresponds to the cavity photons with complex frequency $\Tilde{\omega}_m$ and with creation (annihilation) operator $\hat{\alpha}_{m}^\dag$ ($\hat{\alpha}_{m}$). The second term of equation (\ref{Hfree}) describes the magnon modes as a sum of harmonic oscillators with complex frequency $\Tilde{\omega}_K$ and creation (annihilation) operator $\hat{m}_{l}^\dag$ ($\hat{m}_{l}$). For both photonic and magnetic term, we took into account the corresponding intrinsic damping rates: $\kappa_c$ and $\kappa_m$ such as: $\Tilde{\omega}_m=\omega_m - i\kappa_c/2$ and $\Tilde{\omega}_K=\omega_K-i\kappa_m/2$ respectively, where $\omega_m$ and $\omega_K$ correspond to the eigenfrequencies of the photonic and magnetic modes respectively. On the other hand, the interaction term $\hat{H}_{int}$ of the Hamiltonian, originating from the Zeeman interaction is written as \cite{Yu2020}:

\begin{equation}
\hat{H}_{int}=\hbar\sum_m\sum_{l=1}^N|g_{l,m}|e^{i(sm_*\phi+\theta_s)}\hat{m}_{l}^{\dag}\hat{\alpha}_{m} + H.c,
\label{Hint}
\end{equation}

where $H.c$ stands for Hermitian conjugate. A phase term $\theta_s$ carried by the exponential function is added to the azimuthal coordinate $\phi$, to account for the presence of the specific coupling phase of our system. Such terms have recently been described in cavity magnonics systems \cite{Gardin2023, Gardin2024}, exhibiting behavior analogous to a discrete Pancharatnam–Berry phase \cite{Resta2000}. The coupling phase of our system is defined according to the two polar components of the normalized magnetic field inside the cavity: $\theta_s=\atantwo(s \frac{\mathcal{H}^m_{\phi}}{\mathcal{H}^m_{\rho}})$ where  $\mathcal{H}^m_{\rho}$ and $\mathcal{H}^m_{\phi}$ are the normalized magnetic fields inside the resonator \cite{Yu2020} (see Appendix \hyperref[appendixB1]{B1}). Figure \ref{FIG_2} \textbf{a)} is a mapping of the chiral resonator showing the values of the coupling phases $\theta_s$ with $s$=sgn$(m)$ on the two chiral lines $\rho_+$ and $\rho_-$ for $4$ different values of the azimuthal coordinate: $\phi=\{0,\frac{\pi}{2},\pi,\frac{3\pi}{2}\}$. On the outskirts of the outer circle are displayed the values of the $\phi$-coordinates. The coupling strengths $g_{l,m}$ in equation (\ref{Hint}) read:

\begin{equation}
g_{l,m}=\mu_0 \sqrt{\frac{\gamma M_{s} V_l}{2 \hbar}}(\mathcal{H}^m_{\rho}+is\mathcal{H}^m_{\phi}),
\label{g}
\end{equation}

where $\mu\textsubscript{0}$ and the integer $m$ are vacuum permeability and orbital angular momentum of the cavity mode respectively, $M_s$ the saturation magnetization, $V_l$ the volume of the sphere and $s$=sgn$(m)$, $m_*=|m|+1$. Figure \ref{FIG_1} \textbf{d)} underlines the radial dependency of the coupling strength for $N=1$ and finds good agreement between theory, finite element method (FEM) simulations and measurements. The full derivation of $\hat{H}_{int}$ is detailed in Appendix \hyperref[appendixB1]{B1}.

In order to describe our CMP spectra with multiple identical YIG spheres accurately, we follow a semiclassical approach using standard Input-Output theory within the rotating-wave approximation (RWA). It is worth noting that to give the most complete description of our system, we take into consideration the phase contribution of each external coupling between the cavity and the ports \cite{Bourcin2024}. These contributions depend on the phase of the injected electromagnetic field inside our system. In the case of a single YIG sphere located on the $\rho_-$ chiral line coupling to the CCW mode, the transmission parameter $S_{21}$ reads:

\begin{widetext}
\begin{equation}
    S_{21} = -i \frac{\kappa_{cw2}[\kappa_{ccw1}^*(\frac{G_+^*G_-}{\omega - \omega_K}-\frac{i}{2}J^*) + \kappa_{cw1}^*\Gamma_-] + \kappa_{ccw2}[\kappa_{cw1}^*(\frac{G_+G_-^*}{\omega - \omega_K}-\frac{i}{2}J) + \kappa_{ccw1}^*\Gamma_+]}{\Gamma_-\Gamma_+ - (\frac{G_+G_-^*}{\omega - \omega_K}-\frac{i}{2}J) (\frac{G_+^*G_-}{\omega - \omega_K}-\frac{i}{2}J^*)},
    \label{S21}
\end{equation}
\end{widetext}

where $\kappa_{cw1,ccw1}=\gamma_{cw1,ccw1}e^{i\varphi_{cw1,ccw1}}$, $\kappa_{cw2,ccw2}=\gamma_{cw2,ccw2}e^{i\varphi_{cw2,ccw2}}$ are the terms taking into account both dissipation ratio amplitude ($\gamma_{cw,ccw}$) and related phase ($\varphi_{cw,ccw}$), $G_{\pm}=|g_{l,\pm m}|e^{i(s m_*\phi+\theta_s)}$,  $\Gamma_{\pm}=\omega - \Tilde{\omega}_{\pm} - \frac{|G_{\pm}|^2}{\omega - \Tilde{\omega}_K}$. The $S_{21}$ parameter is plotted for a YIG sphere located at $(\rho_-, 0)$ and $(\rho_+, 0)$ in Figure \ref{FIG_2} \textbf{b)} and \textbf{c)} respectively. An additional term $J$ appears in the expression of $S_{21}$ this term can be explained as an indirect coupling between the two photonic modes, mediated by the two probes, that splits the cavity mode into a doublet \cite{Bourhill2023}. The expression of $J$ can be found in Appendix \hyperref[appendixB2]{B2} as well as the derivation of the transmission parameter for one and two YIG spheres loaded in the cavity. The degeneracy of our system is reflected in equation (\ref{S21}) by the cross products of the two complex coupling strengths $G_\pm$ and their squared absolute values carried by the $\Gamma_\pm$ terms. In order to quantify the non-reciprocal features of our system, we compare the transmission amplitudes originating from port 1 and 2 by using the isolation ratio defined as: 

\begin{equation}
    \text{Iso}.=20*\text{log}_{10}(S_{21}/S_{12}).
    \label{iso body}
\end{equation}

This expression allows to clearly evidence the degree of non-reciprocity in our system for $S_{21} \neq S_{12}$. Figure \ref{FIG_2} \textbf{d)} and \textbf{e)} show the theoretical isolation ratio spectrum obtained for a single YIG sphere at positions $\rho_-$ and $\rho_+$, by making a cut in the spectrum, indicated by the black dashed line, we retrieve a theoretical absolute isolation ratio reaching more than $50$ dB as it is shown in Figure \ref{FIG_2} \textbf{f)} and \textbf{g)}.

\section{Strong coupling in chiral cavity}\phantomsection\label{Sec4}

\begin{figure}
\includegraphics[width=\linewidth]{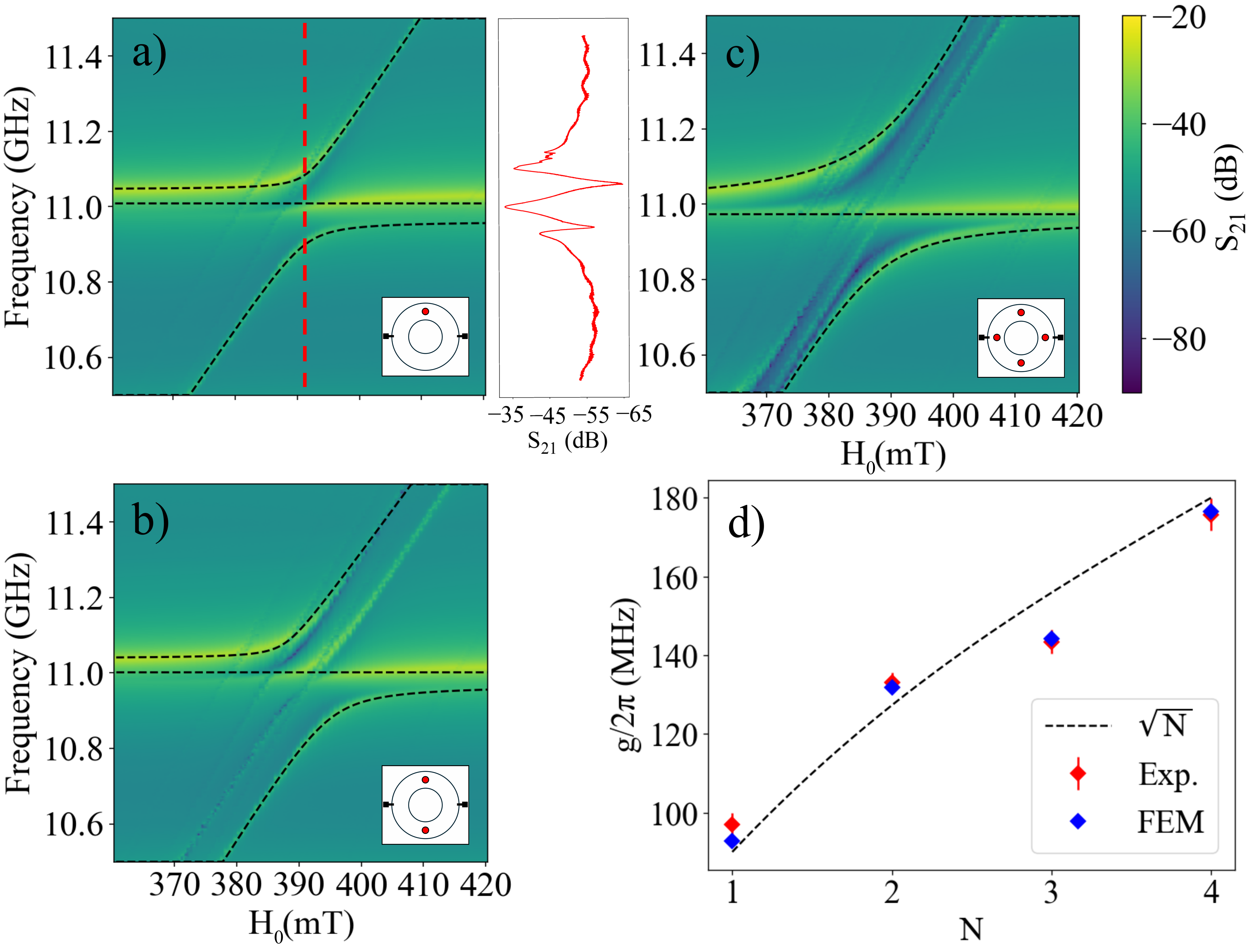}
\caption{The dependency of the coupling strength according to the number of YIG spheres. \textbf{a)}-\textbf{c)} Measured transmission spectra for $N=1,2$ and $4$ YIG spheres respectively, located on the $\rho\textsubscript{-}$ chiral line. The insets denote each configuration of the spheres inside the resonator as well as the configuration of the ports. The black dashed lines correspond to the fitting of the CMP branches using (\ref{S21}). Contiguous to \textbf{a)} is added the transmission parameter $S_{21}$ measured at the field value given by the red dashed curve. The low transmission curves inside the coupling gaps are identified as anti-resonances of the system. \textbf{d)} Simulated (blue) and measured (red) coupling strength $g$ according to the number of YIG spheres $N$ along with the $\sqrt{N}$ curve (black dashed line). The uncertainties attributed to the experimental points are due to the precision of the fitting parameters. The supplementary linear absorption curves before the anticrossing gaps are due to magnetostatic modes inside the YIG spheres \cite{Walker1958}.}
\label{FIG_3}
\end{figure}

\begin{figure*}
\includegraphics[width=0.8\linewidth]{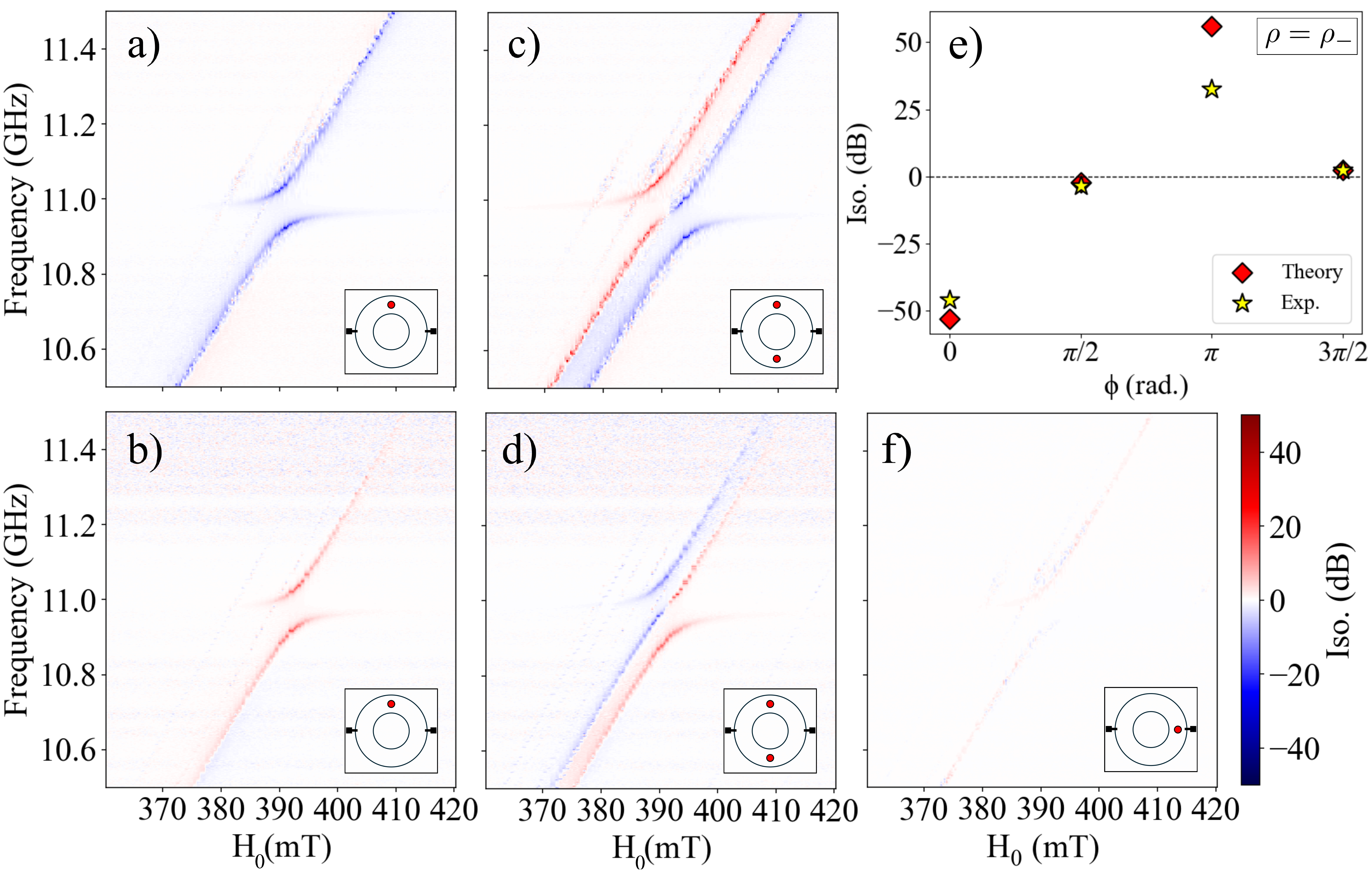}
\caption{Non-reciprocal behavior of the multi-sphere loaded SIW chiral cavity. \textbf{a)} and \textbf{b)} measured Isolation ratio (Iso.) as a function of frequency and bias magnetic field for $N=1$ at coordinates $(\rho_-, 0)$ and $(\rho_+, 0)$ respectively. \textbf{c)}  and \textbf{d)} measured Iso. for $N=2$ at position $(\rho_-, 0)$, $(\rho_-, \pi)$ and $(\rho_+,0)$, $(\rho_+,\pi)$ respectively. \textbf{e)} Dependency of the isolation ratio (Iso.) extrema according to the azimuthal coordinate $\phi$ on the radial line $\rho_-$. \textbf{f)} Measured isolation ratio (Iso.) for $N=1$ at position $(\rho_-,3\pi/2)$. The insets show the number of YIG spheres and their configuration in each case. The isolation ratio for other configurations of multiple YIG spheres are displayed in Appendix \hyperref[appendixC]{C}.}
\label{FIG_4}
\end{figure*}

In this section, we present the results of our transmission measurements for multiple 1 mm diameter YIG spheres loaded in the resonator. Figure \ref{FIG_3} \textbf{a)-c)} shows the measured transmission spectrum for $N=1$, $N=2$ and $N=4$ respectively. In the case where $N=1$, at position $(\rho=\rho_-, \phi=0)$, for which we find good consistency with the theory (see Figure \ref{FIG_2} \textbf{a)}), one can clearly see the cavity mode degeneracy: only one of the two chiral modes is coupled to the Kittel mode of the YIG sphere loaded in the resonator, leaving an almost uncoupled cavity mode (CW) inside the coupling gap. The degeneracy is lifted around the magnon-photon coupling but for higher magnetic field values, the cavity mode will stop appearing as two distinct modes and become degenerate. Figure \ref{FIG_3} \textbf{a)} shows that our interacting photon-magnon system has reached the strong coupling regime, displaying an avoided crossing in transmission measurements. We reach a coupling strength of $97$ MHz which largely exceeds the cavity ($\frac{\kappa_c}{2\pi} = 19.7$ MHz) and magnon ($\frac{\kappa_m}{2\pi} =8$ MHz) linewidth.

The sharp transmission minima (below 60 dB) observed near the coupling gap in Figure \ref{FIG_3} \textbf{a)} to \textbf{c)} are identified as anti-resonances \cite{Harder2016, Bourcin2024}. Those drops in transmission are directly related to the phases associated with the port dissipation ratio that we take into account in the input-output formalism. However, unlike the non-degenerate situation, where the anti-resonance is identified as a single sharp dip in transmission between the two branches \cite{Bourcin2024}, here the anti-resonance is split in two due to the existence of an uncoupled degenerate cavity mode (see Figure \ref{FIG_2} \textbf{a)} and Appendix \hyperref[appendixB2]{B2}). This is the reason why Figure \ref{FIG_3} \textbf{a)}-\textbf{c)} show low transmission curves following the two polaritonic branches. Moreover, by adding more spheres to the $\rho_-$ chiral line, higher order magnetostatic modes appear accordingly. The frequency shift between the linear modes inside the coupling gap is due to the fact that the YIG spheres used in the experiment are not identical and have slightly different resonant frequency from one another. As expected, we observe the typical $\sqrt{N}$ dependence for an increasing number of YIG spheres along the chiral line $\rho_-$ as it is highlighted in Figure \ref{FIG_3} \textbf{d)}. This is because the coupling scales with the number of spins which are allowed to be distributed among several samples when strongly coupled with cavity photons \cite{Yu2020}. Moreover, Figure \ref{FIG_3} \textbf{d)} shows good agreement between FEM simulations and measurements.

\section{\textit{N}-sphere non-reciprocity }\phantomsection\label{Sec5}

Chirality in microwave cavities can be exploited to enable non-reciprocity by selectively coupling cavity photons to magnons with the same polarization \cite{TaoYu2020, Zhang2020}. This section provides the experimental results in terms of non-reciprocity of our system by evaluating the isolation ratio (c.f. equation (\ref{iso body})) between $S_{21}$ and $S_{12}$. We measure an isolation ratio reaching $46$ dB of amplitude for $N=1$ at positions $(\rho_-,0)$, which is in good agreement with the theoretical results displayed in Figure \ref{FIG_2} and twice as high as the isolation ratio measured for toroidal chiral cavities prior to this study. We also find good agreement between theory and experimental data in the case of two YIG spheres loaded in the cavity as it is shown in Appendix \hyperref[appendixB3]{B3} \cite{Bourhill2023, Zhang2020}.

In the case where the YIG sphere is located in the top part of the resonator at radial and azimuthal coordinates $\rho=\rho_-$ and $\phi=0$, Figure \ref{FIG_4} \textbf{a)} shows a dominance of the $S_{12}$ parameter over $S_{21}$ which is reversed if the position of the sphere is rotated by $\pi$ around the $z$-axis (see Appendix \hyperref[appendixC]{C}). This observation highlights a direct dependency of the non-reciprocity along $\phi$. Interestingly, we can find the same feature but reversed if the radial coordinate of the YIG sphere is switched to $\rho_+$, in this case the $S_{21}$ parameter becomes dominant over $S_{12}$ as illustrated in Figure \ref{FIG_4} \textbf{b)}. Figure \ref{FIG_4} \textbf{c)} and \textbf{d)} show that for two YIG spheres at azimuthal positions $\phi=0$ and $\phi=\pi$ both non-reciprocal features of top and bottom YIG sphere combine. Figure \ref{FIG_4} \textbf{c)} shows a red-dominant curve above the condition of resonant coupling between cavity photons and Kittel mode and blue-dominant below on the $\rho_-$ chiral line, and vice versa for $\rho_+$ (Figure \ref{FIG_4} \textbf{d)}). The theoretical plots for the same $N=2$ YIG spheres configuration can be found in Appendix \hyperref[appendixB3]{B3}.

Figure \ref{FIG_4} \textbf{e)} displays the dependency of the isolation ratio extrema according to the $\phi$ coordinate on the $\rho_-$ chiral line obtained for both theoretical and experimental results. One can see that for lateral positions of the YIG sphere, there is very weak non-reciprocity. This observation concurs with the plot of Figure \ref{FIG_4} \textbf{f)} showing the isolation ratio according to DC magnetic field and frequency at position $(\rho_-,3\pi/2)$ for which the system is almost reciprocal. Moreover, it is worth to be noted that if the position of the two YIG spheres described in Figure \ref{FIG_4} \textbf{c)} and \textbf{d)} is rotated by an angle of 90° we obtain again an important attenuation of non-reciprocity (see Appendix \hyperref[appendixC]{C}) by combination of reciprocal positions. This amplification or cancellation of non-reciprocity is also connected to the direction of the energy flow as it was developed in \cite{Bourhill2023}. Those observations demonstrate that we can fully control the isolation ratio, and ultimately the non-reciprocity, by tuning the positions of the YIG spheres inside the resonator and thus the degree of non-reciprocity in our system at will. By comparing Figure \ref{FIG_3} and Figure \ref{FIG_4}, one can see in the spectra that the isolation ratio is nonzero exactly where the anti-resonances are located. This observation confirms that the anti-resonances highlighted in Figure \ref{FIG_3} are effectively the source of non-reciprocal behaviors in our system \cite{Bourhill2023,Bourcin2024}.

By taking a closer look at Figure \ref{FIG_2} \textbf{a)} that at positions $(\rho_-,\frac{\pi}{2})$ and $(\rho_-,\frac{3\pi}{2})$ the coupling phases are in phase opposition, where at position $(\rho_-,0)$ and $(\rho_-,\pi)$ they share the same value. This observation provides insight on how the coupling phase associated with each sphere is linked to the effective non-reciprocity observed in the system. The coupling phase difference, arising from the spatial dependence imposed by the system’s chirality, results in either an enhancement or suppression of the observed non-reciprocity. Adding more spheres inside the cavity ($N=3$, $N=4$) deeply changes the shape and intensity of the measured isolation ratio (see Appendix \hyperref[appendixC]{C}) because the addition of YIG spheres leads also to an addition of coupling phase.

\section{Conclusion}\phantomsection\label{Sec6}

In this study, we explored chirality-based non-reciprocal transmission in a toroidal cavity optimized through SIW technology. The chirality of our resonator arises from the spin angular momentum of the microwave photons. This property enables the existence of hybridization selectivity with magnetic samples at specific chiral lines ($\rho_-$ and $\rho_+$). By adding $N$ spheres of YIG in our resonant system, we validated the $\sqrt{N}$ dependency of the coupling strength predicted by \cite{Yu2020}, this finding shows strong consistency between FEM simulations and experimental results. Moreover, we developed a comprehensive theoretical model based on the input-output formalism to explain the behavior of non-reciprocity according to the number of YIG spheres and find good agreement between analytical theory, simulations and measurements. In this study, we introduced the coupling phase of strongly coupled magnon-photon system \cite{Gardin2023} which had not been explicitly defined in earlier works on chiral cavity magnonics \cite{Yu2020, Bourhill2023}. We relate this phase term to the emergence of amplified or attenuated non-reciprocal effects, providing full control over the system's behavior. The design of our cavity provides full control over the isolation ratio depending on the disposition of the magnetic samples, effectively enabling or disabling non-reciprocity on demand.

Our contribution sheds a new light on the mechanisms and tunability of non-reciprocity, laying the groundwork for further investigation of multi-sphere cavity magnonics. Thus, this work offers promising prospects for an emerging class of non-reciprocal devices that could also be used for cavity-mediated magnon-magnon interaction \cite{Zhang2015}, paving the way for the development of multimode quantum memories.

\begin{acknowledgments}
We acknowledge the financial support of ANR project ICARUS under grant agreement: ANR-22-CE24-0008-03. ICARUS is a collaborative project between IMT Atlantique, Thales and Elliptika. We acknowledge Bernard Abiven as well for CNC machining.
\end{acknowledgments}

\nocite{*}

\bibliographystyle{apsrev4-1}
\bibliography{apssamp}

\begin{thebibliography}{42}%
\makeatletter
\providecommand \@ifxundefined [1]{%
 \@ifx{#1\undefined}
}%
\providecommand \@ifnum [1]{%
 \ifnum #1\expandafter \@firstoftwo
 \else \expandafter \@secondoftwo
 \fi
}%
\providecommand \@ifx [1]{%
 \ifx #1\expandafter \@firstoftwo
 \else \expandafter \@secondoftwo
 \fi
}%
\providecommand \natexlab [1]{#1}%
\providecommand \enquote  [1]{``#1''}%
\providecommand \bibnamefont  [1]{#1}%
\providecommand \bibfnamefont [1]{#1}%
\providecommand \citenamefont [1]{#1}%
\providecommand \href@noop [0]{\@secondoftwo}%
\providecommand \href [0]{\begingroup \@sanitize@url \@href}%
\providecommand \@href[1]{\@@startlink{#1}\@@href}%
\providecommand \@@href[1]{\endgroup#1\@@endlink}%
\providecommand \@sanitize@url [0]{\catcode `\\12\catcode `\$12\catcode `\&12\catcode `\#12\catcode `\^12\catcode `\_12\catcode `\%12\relax}%
\providecommand \@@startlink[1]{}%
\providecommand \@@endlink[0]{}%
\providecommand \url  [0]{\begingroup\@sanitize@url \@url }%
\providecommand \@url [1]{\endgroup\@href {#1}{\urlprefix }}%
\providecommand \urlprefix  [0]{URL }%
\providecommand \Eprint [0]{\href }%
\providecommand \doibase [0]{http://dx.doi.org/}%
\providecommand \selectlanguage [0]{\@gobble}%
\providecommand \bibinfo  [0]{\@secondoftwo}%
\providecommand \bibfield  [0]{\@secondoftwo}%
\providecommand \translation [1]{[#1]}%
\providecommand \BibitemOpen [0]{}%
\providecommand \bibitemStop [0]{}%
\providecommand \bibitemNoStop [0]{.\EOS\space}%
\providecommand \EOS [0]{\spacefactor3000\relax}%
\providecommand \BibitemShut  [1]{\csname bibitem#1\endcsname}%
\let\auto@bib@innerbib\@empty
\bibitem [{\citenamefont {Huebl}\ \emph {et~al.}(2013)\citenamefont {Huebl}, \citenamefont {Zollitsch}, \citenamefont {Lotze}, \citenamefont {Hocke}, \citenamefont {Greifenstein}, \citenamefont {Marx}, \citenamefont {Gross},\ and\ \citenamefont {Goennenwein}}]{Huebl2013}%
  \BibitemOpen
  \bibfield  {author} {\bibinfo {author} {\bibfnamefont {H.}~\bibnamefont {Huebl}}, \bibinfo {author} {\bibfnamefont {C.~W.}\ \bibnamefont {Zollitsch}}, \bibinfo {author} {\bibfnamefont {J.}~\bibnamefont {Lotze}}, \bibinfo {author} {\bibfnamefont {F.}~\bibnamefont {Hocke}}, \bibinfo {author} {\bibfnamefont {M.}~\bibnamefont {Greifenstein}}, \bibinfo {author} {\bibfnamefont {A.}~\bibnamefont {Marx}}, \bibinfo {author} {\bibfnamefont {R.}~\bibnamefont {Gross}}, \ and\ \bibinfo {author} {\bibfnamefont {S.~T.~B.}\ \bibnamefont {Goennenwein}},\ }\href {\doibase 10.1103/physrevlett.111.127003} {\bibfield  {journal} {\bibinfo  {journal} {Physical Review Letters}\ }\textbf {\bibinfo {volume} {111}} (\bibinfo {year} {2013}),\ 10.1103/physrevlett.111.127003}\BibitemShut {NoStop}%
\bibitem [{\citenamefont {Tabuchi}\ \emph {et~al.}(2014)\citenamefont {Tabuchi}, \citenamefont {Ishino}, \citenamefont {Ishikawa}, \citenamefont {Yamazaki}, \citenamefont {Usami},\ and\ \citenamefont {Nakamura}}]{Tabuchi2014}%
  \BibitemOpen
  \bibfield  {author} {\bibinfo {author} {\bibfnamefont {Y.}~\bibnamefont {Tabuchi}}, \bibinfo {author} {\bibfnamefont {S.}~\bibnamefont {Ishino}}, \bibinfo {author} {\bibfnamefont {T.}~\bibnamefont {Ishikawa}}, \bibinfo {author} {\bibfnamefont {R.}~\bibnamefont {Yamazaki}}, \bibinfo {author} {\bibfnamefont {K.}~\bibnamefont {Usami}}, \ and\ \bibinfo {author} {\bibfnamefont {Y.}~\bibnamefont {Nakamura}},\ }\href {\doibase 10.1103/physrevlett.113.083603} {\bibfield  {journal} {\bibinfo  {journal} {Physical Review Letters}\ }\textbf {\bibinfo {volume} {113}} (\bibinfo {year} {2014}),\ 10.1103/physrevlett.113.083603}\BibitemShut {NoStop}%
\bibitem [{\citenamefont {Zhang}\ \emph {et~al.}(2014)\citenamefont {Zhang}, \citenamefont {Zou}, \citenamefont {Jiang},\ and\ \citenamefont {Tang}}]{Zhang2014}%
  \BibitemOpen
  \bibfield  {author} {\bibinfo {author} {\bibfnamefont {X.}~\bibnamefont {Zhang}}, \bibinfo {author} {\bibfnamefont {C.-L.}\ \bibnamefont {Zou}}, \bibinfo {author} {\bibfnamefont {L.}~\bibnamefont {Jiang}}, \ and\ \bibinfo {author} {\bibfnamefont {H.~X.}\ \bibnamefont {Tang}},\ }\href {\doibase 10.1103/physrevlett.113.156401} {\bibfield  {journal} {\bibinfo  {journal} {Physical Review Letters}\ }\textbf {\bibinfo {volume} {113}} (\bibinfo {year} {2014}),\ 10.1103/physrevlett.113.156401}\BibitemShut {NoStop}%
\bibitem [{\citenamefont {Soykal}\ and\ \citenamefont {Flatté}(2010)}]{Soykal2010}%
  \BibitemOpen
  \bibfield  {author} {\bibinfo {author} {\bibfnamefont {O.~O.}\ \bibnamefont {Soykal}}\ and\ \bibinfo {author} {\bibfnamefont {M.~E.}\ \bibnamefont {Flatté}},\ }\href {\doibase 10.1103/physrevlett.104.077202} {\bibfield  {journal} {\bibinfo  {journal} {Physical Review Letters}\ }\textbf {\bibinfo {volume} {104}} (\bibinfo {year} {2010}),\ 10.1103/physrevlett.104.077202}\BibitemShut {NoStop}%
\bibitem [{\citenamefont {Flower}\ \emph {et~al.}(2019)\citenamefont {Flower}, \citenamefont {Goryachev}, \citenamefont {Bourhill},\ and\ \citenamefont {Tobar}}]{Flower2019}%
  \BibitemOpen
  \bibfield  {author} {\bibinfo {author} {\bibfnamefont {G.}~\bibnamefont {Flower}}, \bibinfo {author} {\bibfnamefont {M.}~\bibnamefont {Goryachev}}, \bibinfo {author} {\bibfnamefont {J.}~\bibnamefont {Bourhill}}, \ and\ \bibinfo {author} {\bibfnamefont {M.~E.}\ \bibnamefont {Tobar}},\ }\href {\doibase 10.1088/1367-2630/ab3e1c} {\bibfield  {journal} {\bibinfo  {journal} {New Journal of Physics}\ }\textbf {\bibinfo {volume} {21}},\ \bibinfo {pages} {095004} (\bibinfo {year} {2019})}\BibitemShut {NoStop}%
\bibitem [{\citenamefont {Bourcin}\ \emph {et~al.}(2023)\citenamefont {Bourcin}, \citenamefont {Bourhill}, \citenamefont {Vlaminck},\ and\ \citenamefont {Castel}}]{Bourcin2023}%
  \BibitemOpen
  \bibfield  {author} {\bibinfo {author} {\bibfnamefont {G.}~\bibnamefont {Bourcin}}, \bibinfo {author} {\bibfnamefont {J.}~\bibnamefont {Bourhill}}, \bibinfo {author} {\bibfnamefont {V.}~\bibnamefont {Vlaminck}}, \ and\ \bibinfo {author} {\bibfnamefont {V.}~\bibnamefont {Castel}},\ }\href {\doibase 10.1103/physrevb.107.214423} {\bibfield  {journal} {\bibinfo  {journal} {Physical Review B}\ }\textbf {\bibinfo {volume} {107}} (\bibinfo {year} {2023}),\ 10.1103/physrevb.107.214423}\BibitemShut {NoStop}%
\bibitem [{\citenamefont {Harder}\ \emph {et~al.}(2018)\citenamefont {Harder}, \citenamefont {Yang}, \citenamefont {Yao}, \citenamefont {Yu}, \citenamefont {Rao}, \citenamefont {Gui}, \citenamefont {Stamps},\ and\ \citenamefont {Hu}}]{Harder2018}%
  \BibitemOpen
  \bibfield  {author} {\bibinfo {author} {\bibfnamefont {M.}~\bibnamefont {Harder}}, \bibinfo {author} {\bibfnamefont {Y.}~\bibnamefont {Yang}}, \bibinfo {author} {\bibfnamefont {B.}~\bibnamefont {Yao}}, \bibinfo {author} {\bibfnamefont {C.}~\bibnamefont {Yu}}, \bibinfo {author} {\bibfnamefont {J.}~\bibnamefont {Rao}}, \bibinfo {author} {\bibfnamefont {Y.}~\bibnamefont {Gui}}, \bibinfo {author} {\bibfnamefont {R.}~\bibnamefont {Stamps}}, \ and\ \bibinfo {author} {\bibfnamefont {C.-M.}\ \bibnamefont {Hu}},\ }\href {\doibase 10.1103/physrevlett.121.137203} {\bibfield  {journal} {\bibinfo  {journal} {Physical Review Letters}\ }\textbf {\bibinfo {volume} {121}} (\bibinfo {year} {2018}),\ 10.1103/physrevlett.121.137203}\BibitemShut {NoStop}%
\bibitem [{\citenamefont {Zhang}\ \emph {et~al.}(2017)\citenamefont {Zhang}, \citenamefont {Luo}, \citenamefont {Wang}, \citenamefont {Li},\ and\ \citenamefont {You}}]{Zhang2017}%
  \BibitemOpen
  \bibfield  {author} {\bibinfo {author} {\bibfnamefont {D.}~\bibnamefont {Zhang}}, \bibinfo {author} {\bibfnamefont {X.-Q.}\ \bibnamefont {Luo}}, \bibinfo {author} {\bibfnamefont {Y.-P.}\ \bibnamefont {Wang}}, \bibinfo {author} {\bibfnamefont {T.-F.}\ \bibnamefont {Li}}, \ and\ \bibinfo {author} {\bibfnamefont {J.~Q.}\ \bibnamefont {You}},\ }\href {\doibase 10.1038/s41467-017-01634-w} {\bibfield  {journal} {\bibinfo  {journal} {Nature Communications}\ }\textbf {\bibinfo {volume} {8}} (\bibinfo {year} {2017}),\ 10.1038/s41467-017-01634-w}\BibitemShut {NoStop}%
\bibitem [{\citenamefont {Rao}\ \emph {et~al.}(2021)\citenamefont {Rao}, \citenamefont {Xu}, \citenamefont {Gui}, \citenamefont {Wang}, \citenamefont {Yang}, \citenamefont {Yao}, \citenamefont {Dietrich}, \citenamefont {Bridges}, \citenamefont {Fan}, \citenamefont {Xue},\ and\ \citenamefont {Hu}}]{Rao2021}%
  \BibitemOpen
  \bibfield  {author} {\bibinfo {author} {\bibfnamefont {J.~W.}\ \bibnamefont {Rao}}, \bibinfo {author} {\bibfnamefont {P.~C.}\ \bibnamefont {Xu}}, \bibinfo {author} {\bibfnamefont {Y.~S.}\ \bibnamefont {Gui}}, \bibinfo {author} {\bibfnamefont {Y.~P.}\ \bibnamefont {Wang}}, \bibinfo {author} {\bibfnamefont {Y.}~\bibnamefont {Yang}}, \bibinfo {author} {\bibfnamefont {B.}~\bibnamefont {Yao}}, \bibinfo {author} {\bibfnamefont {J.}~\bibnamefont {Dietrich}}, \bibinfo {author} {\bibfnamefont {G.~E.}\ \bibnamefont {Bridges}}, \bibinfo {author} {\bibfnamefont {X.~L.}\ \bibnamefont {Fan}}, \bibinfo {author} {\bibfnamefont {D.~S.}\ \bibnamefont {Xue}}, \ and\ \bibinfo {author} {\bibfnamefont {C.-M.}\ \bibnamefont {Hu}},\ }\href {\doibase 10.1038/s41467-021-22171-7} {\bibfield  {journal} {\bibinfo  {journal} {Nature Communications}\ }\textbf {\bibinfo {volume} {12}} (\bibinfo {year} {2021}),\ 10.1038/s41467-021-22171-7}\BibitemShut {NoStop}%
\bibitem [{\citenamefont {Lambert}\ \emph {et~al.}(2016)\citenamefont {Lambert}, \citenamefont {Haigh}, \citenamefont {Langenfeld}, \citenamefont {Doherty},\ and\ \citenamefont {Ferguson}}]{Lambert2016}%
  \BibitemOpen
  \bibfield  {author} {\bibinfo {author} {\bibfnamefont {N.~J.}\ \bibnamefont {Lambert}}, \bibinfo {author} {\bibfnamefont {J.~A.}\ \bibnamefont {Haigh}}, \bibinfo {author} {\bibfnamefont {S.}~\bibnamefont {Langenfeld}}, \bibinfo {author} {\bibfnamefont {A.~C.}\ \bibnamefont {Doherty}}, \ and\ \bibinfo {author} {\bibfnamefont {A.~J.}\ \bibnamefont {Ferguson}},\ }\href {\doibase 10.1103/physreva.93.021803} {\bibfield  {journal} {\bibinfo  {journal} {Physical Review A}\ }\textbf {\bibinfo {volume} {93}} (\bibinfo {year} {2016}),\ 10.1103/physreva.93.021803}\BibitemShut {NoStop}%
\bibitem [{\citenamefont {Zare~Rameshti}\ and\ \citenamefont {Bauer}(2018)}]{ZareRameshti2018}%
  \BibitemOpen
  \bibfield  {author} {\bibinfo {author} {\bibfnamefont {B.}~\bibnamefont {Zare~Rameshti}}\ and\ \bibinfo {author} {\bibfnamefont {G.~E.~W.}\ \bibnamefont {Bauer}},\ }\href {\doibase 10.1103/physrevb.97.014419} {\bibfield  {journal} {\bibinfo  {journal} {Physical Review B}\ }\textbf {\bibinfo {volume} {97}} (\bibinfo {year} {2018}),\ 10.1103/physrevb.97.014419}\BibitemShut {NoStop}%
\bibitem [{\citenamefont {Wang}\ \emph {et~al.}(2019)\citenamefont {Wang}, \citenamefont {Rao}, \citenamefont {Yang}, \citenamefont {Xu}, \citenamefont {Gui}, \citenamefont {Yao}, \citenamefont {You},\ and\ \citenamefont {Hu}}]{Wang2019}%
  \BibitemOpen
  \bibfield  {author} {\bibinfo {author} {\bibfnamefont {Y.-P.}\ \bibnamefont {Wang}}, \bibinfo {author} {\bibfnamefont {J.}~\bibnamefont {Rao}}, \bibinfo {author} {\bibfnamefont {Y.}~\bibnamefont {Yang}}, \bibinfo {author} {\bibfnamefont {P.-C.}\ \bibnamefont {Xu}}, \bibinfo {author} {\bibfnamefont {Y.}~\bibnamefont {Gui}}, \bibinfo {author} {\bibfnamefont {B.}~\bibnamefont {Yao}}, \bibinfo {author} {\bibfnamefont {J.}~\bibnamefont {You}}, \ and\ \bibinfo {author} {\bibfnamefont {C.-M.}\ \bibnamefont {Hu}},\ }\href {\doibase 10.1103/physrevlett.123.127202} {\bibfield  {journal} {\bibinfo  {journal} {Physical Review Letters}\ }\textbf {\bibinfo {volume} {123}} (\bibinfo {year} {2019}),\ 10.1103/physrevlett.123.127202}\BibitemShut {NoStop}%
\bibitem [{\citenamefont {Zhang}\ \emph {et~al.}(2020)\citenamefont {Zhang}, \citenamefont {Galda}, \citenamefont {Han}, \citenamefont {Jin},\ and\ \citenamefont {Vinokur}}]{Zhang2020}%
  \BibitemOpen
  \bibfield  {author} {\bibinfo {author} {\bibfnamefont {X.}~\bibnamefont {Zhang}}, \bibinfo {author} {\bibfnamefont {A.}~\bibnamefont {Galda}}, \bibinfo {author} {\bibfnamefont {X.}~\bibnamefont {Han}}, \bibinfo {author} {\bibfnamefont {D.}~\bibnamefont {Jin}}, \ and\ \bibinfo {author} {\bibfnamefont {V.~M.}\ \bibnamefont {Vinokur}},\ }\href {\doibase 10.1103/physrevapplied.13.044039} {\bibfield  {journal} {\bibinfo  {journal} {Physical Review Applied}\ }\textbf {\bibinfo {volume} {13}} (\bibinfo {year} {2020}),\ 10.1103/physrevapplied.13.044039}\BibitemShut {NoStop}%
\bibitem [{\citenamefont {Zhong}\ \emph {et~al.}(2022)\citenamefont {Zhong}, \citenamefont {Zhang},\ and\ \citenamefont {Yao}}]{Zhong2022}%
  \BibitemOpen
  \bibfield  {author} {\bibinfo {author} {\bibfnamefont {L.}~\bibnamefont {Zhong}}, \bibinfo {author} {\bibfnamefont {C.}~\bibnamefont {Zhang}}, \ and\ \bibinfo {author} {\bibfnamefont {B.~M.}\ \bibnamefont {Yao}},\ }\href {\doibase 10.1063/5.0102155} {\bibfield  {journal} {\bibinfo  {journal} {AIP Advances}\ }\textbf {\bibinfo {volume} {12}} (\bibinfo {year} {2022}),\ 10.1063/5.0102155}\BibitemShut {NoStop}%
\bibitem [{\citenamefont {Lodahl}\ \emph {et~al.}(2017)\citenamefont {Lodahl}, \citenamefont {Mahmoodian}, \citenamefont {Stobbe}, \citenamefont {Rauschenbeutel}, \citenamefont {Schneeweiss}, \citenamefont {Volz}, \citenamefont {Pichler},\ and\ \citenamefont {Zoller}}]{Lodahl2017}%
  \BibitemOpen
  \bibfield  {author} {\bibinfo {author} {\bibfnamefont {P.}~\bibnamefont {Lodahl}}, \bibinfo {author} {\bibfnamefont {S.}~\bibnamefont {Mahmoodian}}, \bibinfo {author} {\bibfnamefont {S.}~\bibnamefont {Stobbe}}, \bibinfo {author} {\bibfnamefont {A.}~\bibnamefont {Rauschenbeutel}}, \bibinfo {author} {\bibfnamefont {P.}~\bibnamefont {Schneeweiss}}, \bibinfo {author} {\bibfnamefont {J.}~\bibnamefont {Volz}}, \bibinfo {author} {\bibfnamefont {H.}~\bibnamefont {Pichler}}, \ and\ \bibinfo {author} {\bibfnamefont {P.}~\bibnamefont {Zoller}},\ }\href {\doibase 10.1038/nature21037} {\bibfield  {journal} {\bibinfo  {journal} {Nature}\ }\textbf {\bibinfo {volume} {541}},\ \bibinfo {pages} {473–480} (\bibinfo {year} {2017})}\BibitemShut {NoStop}%
\bibitem [{\citenamefont {Coles}\ \emph {et~al.}(2016)\citenamefont {Coles}, \citenamefont {Price}, \citenamefont {Dixon}, \citenamefont {Royall}, \citenamefont {Clarke}, \citenamefont {Kok}, \citenamefont {Skolnick}, \citenamefont {Fox},\ and\ \citenamefont {Makhonin}}]{Coles2016}%
  \BibitemOpen
  \bibfield  {author} {\bibinfo {author} {\bibfnamefont {R.~J.}\ \bibnamefont {Coles}}, \bibinfo {author} {\bibfnamefont {D.~M.}\ \bibnamefont {Price}}, \bibinfo {author} {\bibfnamefont {J.~E.}\ \bibnamefont {Dixon}}, \bibinfo {author} {\bibfnamefont {B.}~\bibnamefont {Royall}}, \bibinfo {author} {\bibfnamefont {E.}~\bibnamefont {Clarke}}, \bibinfo {author} {\bibfnamefont {P.}~\bibnamefont {Kok}}, \bibinfo {author} {\bibfnamefont {M.~S.}\ \bibnamefont {Skolnick}}, \bibinfo {author} {\bibfnamefont {A.~M.}\ \bibnamefont {Fox}}, \ and\ \bibinfo {author} {\bibfnamefont {M.~N.}\ \bibnamefont {Makhonin}},\ }\href {\doibase 10.1038/ncomms11183} {\bibfield  {journal} {\bibinfo  {journal} {Nature Communications}\ }\textbf {\bibinfo {volume} {7}} (\bibinfo {year} {2016}),\ 10.1038/ncomms11183}\BibitemShut {NoStop}%
\bibitem [{\citenamefont {Zhang}\ \emph {et~al.}(2011)\citenamefont {Zhang}, \citenamefont {Wei}, \citenamefont {Bao}, \citenamefont {Håkanson}, \citenamefont {Halas}, \citenamefont {Nordlander},\ and\ \citenamefont {Xu}}]{Zhang2011}%
  \BibitemOpen
  \bibfield  {author} {\bibinfo {author} {\bibfnamefont {S.}~\bibnamefont {Zhang}}, \bibinfo {author} {\bibfnamefont {H.}~\bibnamefont {Wei}}, \bibinfo {author} {\bibfnamefont {K.}~\bibnamefont {Bao}}, \bibinfo {author} {\bibfnamefont {U.}~\bibnamefont {Håkanson}}, \bibinfo {author} {\bibfnamefont {N.~J.}\ \bibnamefont {Halas}}, \bibinfo {author} {\bibfnamefont {P.}~\bibnamefont {Nordlander}}, \ and\ \bibinfo {author} {\bibfnamefont {H.}~\bibnamefont {Xu}},\ }\href {\doibase 10.1103/physrevlett.107.096801} {\bibfield  {journal} {\bibinfo  {journal} {Physical Review Letters}\ }\textbf {\bibinfo {volume} {107}} (\bibinfo {year} {2011}),\ 10.1103/physrevlett.107.096801}\BibitemShut {NoStop}%
\bibitem [{\citenamefont {Zaz}\ \emph {et~al.}(2025)\citenamefont {Zaz}, \citenamefont {Chin}, \citenamefont {Viswan}, \citenamefont {Subedi}, \citenamefont {Mishra}, \citenamefont {McElveen}, \citenamefont {Tamang}, \citenamefont {Shapiro}, \citenamefont {N’Diaye}, \citenamefont {Lai},\ and\ \citenamefont {Dowben}}]{Zaz2025}%
  \BibitemOpen
  \bibfield  {author} {\bibinfo {author} {\bibfnamefont {M.~Z.}\ \bibnamefont {Zaz}}, \bibinfo {author} {\bibfnamefont {W.~K.}\ \bibnamefont {Chin}}, \bibinfo {author} {\bibfnamefont {G.}~\bibnamefont {Viswan}}, \bibinfo {author} {\bibfnamefont {A.}~\bibnamefont {Subedi}}, \bibinfo {author} {\bibfnamefont {E.}~\bibnamefont {Mishra}}, \bibinfo {author} {\bibfnamefont {K.~A.}\ \bibnamefont {McElveen}}, \bibinfo {author} {\bibfnamefont {B.}~\bibnamefont {Tamang}}, \bibinfo {author} {\bibfnamefont {D.}~\bibnamefont {Shapiro}}, \bibinfo {author} {\bibfnamefont {A.~T.}\ \bibnamefont {N’Diaye}}, \bibinfo {author} {\bibfnamefont {R.~Y.}\ \bibnamefont {Lai}}, \ and\ \bibinfo {author} {\bibfnamefont {P.~A.}\ \bibnamefont {Dowben}},\ }\href {\doibase 10.1088/1361-648x/ada338} {\bibfield  {journal} {\bibinfo  {journal} {Journal of Physics: Condensed Matter}\ }\textbf {\bibinfo {volume} {37}},\ \bibinfo {pages} {10LT01} (\bibinfo {year} {2025})}\BibitemShut {NoStop}%
\bibitem [{\citenamefont {Litvinov}(2024)}]{Litvinov2024}%
  \BibitemOpen
  \bibfield  {author} {\bibinfo {author} {\bibfnamefont {V.~I.}\ \bibnamefont {Litvinov}},\ }\href {\doibase 10.1088/1361-648x/ad8f83} {\bibfield  {journal} {\bibinfo  {journal} {Journal of Physics: Condensed Matter}\ }\textbf {\bibinfo {volume} {37}},\ \bibinfo {pages} {055001} (\bibinfo {year} {2024})}\BibitemShut {NoStop}%
\bibitem [{\citenamefont {Yu}\ \emph {et~al.}(2020{\natexlab{a}})\citenamefont {Yu}, \citenamefont {Yu},\ and\ \citenamefont {Bauer}}]{Yu2020}%
  \BibitemOpen
  \bibfield  {author} {\bibinfo {author} {\bibfnamefont {W.}~\bibnamefont {Yu}}, \bibinfo {author} {\bibfnamefont {T.}~\bibnamefont {Yu}}, \ and\ \bibinfo {author} {\bibfnamefont {G.~E.~W.}\ \bibnamefont {Bauer}},\ }\href {\doibase 10.1103/physrevb.102.064416} {\bibfield  {journal} {\bibinfo  {journal} {Physical Review B}\ }\textbf {\bibinfo {volume} {102}} (\bibinfo {year} {2020}{\natexlab{a}}),\ 10.1103/physrevb.102.064416}\BibitemShut {NoStop}%
\bibitem [{\citenamefont {Bliokh}\ \emph {et~al.}(2017)\citenamefont {Bliokh}, \citenamefont {Bekshaev},\ and\ \citenamefont {Nori}}]{Bliokh2017}%
  \BibitemOpen
  \bibfield  {author} {\bibinfo {author} {\bibfnamefont {K.~Y.}\ \bibnamefont {Bliokh}}, \bibinfo {author} {\bibfnamefont {A.~Y.}\ \bibnamefont {Bekshaev}}, \ and\ \bibinfo {author} {\bibfnamefont {F.}~\bibnamefont {Nori}},\ }\href {\doibase 10.1103/physrevlett.119.073901} {\bibfield  {journal} {\bibinfo  {journal} {Physical Review Letters}\ }\textbf {\bibinfo {volume} {119}} (\bibinfo {year} {2017}),\ 10.1103/physrevlett.119.073901}\BibitemShut {NoStop}%
\bibitem [{\citenamefont {Hou}\ \emph {et~al.}(2025)\citenamefont {Hou}, \citenamefont {Ren}, \citenamefont {Xie}, \citenamefont {Ma}, \citenamefont {Li},\ and\ \citenamefont {Li}}]{Hou2025}%
  \BibitemOpen
  \bibfield  {author} {\bibinfo {author} {\bibfnamefont {Y.}~\bibnamefont {Hou}}, \bibinfo {author} {\bibfnamefont {Y.-L.}\ \bibnamefont {Ren}}, \bibinfo {author} {\bibfnamefont {J.-K.}\ \bibnamefont {Xie}}, \bibinfo {author} {\bibfnamefont {S.}~\bibnamefont {Ma}}, \bibinfo {author} {\bibfnamefont {F.-L.}\ \bibnamefont {Li}}, \ and\ \bibinfo {author} {\bibfnamefont {P.-B.}\ \bibnamefont {Li}},\ }\href {\doibase 10.1209/0295-5075/adab8a} {\bibfield  {journal} {\bibinfo  {journal} {Europhysics Letters}\ } (\bibinfo {year} {2025}),\ 10.1209/0295-5075/adab8a}\BibitemShut {NoStop}%
\bibitem [{\citenamefont {Zhao}\ \emph {et~al.}(2025)\citenamefont {Zhao}, \citenamefont {Yang}, \citenamefont {Wang}, \citenamefont {Chen}, \citenamefont {Song}, \citenamefont {Ma}, \citenamefont {Yue}, \citenamefont {Liu}, \citenamefont {Sun}, \citenamefont {Rao}, \citenamefont {Yao},\ and\ \citenamefont {Lu}}]{Zhao2025}%
  \BibitemOpen
  \bibfield  {author} {\bibinfo {author} {\bibfnamefont {K.}~\bibnamefont {Zhao}}, \bibinfo {author} {\bibfnamefont {F.}~\bibnamefont {Yang}}, \bibinfo {author} {\bibfnamefont {C.}~\bibnamefont {Wang}}, \bibinfo {author} {\bibfnamefont {Z.}~\bibnamefont {Chen}}, \bibinfo {author} {\bibfnamefont {J.}~\bibnamefont {Song}}, \bibinfo {author} {\bibfnamefont {S.}~\bibnamefont {Ma}}, \bibinfo {author} {\bibfnamefont {Z.}~\bibnamefont {Yue}}, \bibinfo {author} {\bibfnamefont {W.}~\bibnamefont {Liu}}, \bibinfo {author} {\bibfnamefont {L.}~\bibnamefont {Sun}}, \bibinfo {author} {\bibfnamefont {J.}~\bibnamefont {Rao}}, \bibinfo {author} {\bibfnamefont {B.}~\bibnamefont {Yao}}, \ and\ \bibinfo {author} {\bibfnamefont {W.}~\bibnamefont {Lu}},\ }\href {\doibase 10.1063/5.0248518} {\bibfield  {journal} {\bibinfo  {journal} {AIP Advances}\ }\textbf {\bibinfo {volume} {15}} (\bibinfo {year} {2025}),\ 10.1063/5.0248518}\BibitemShut {NoStop}%
\bibitem [{\citenamefont {Liu}\ \emph {et~al.}(2024)\citenamefont {Liu}, \citenamefont {Chen},\ and\ \citenamefont {Shao}}]{https://doi.org/10.48550/arxiv.2412.10888}%
  \BibitemOpen
  \bibfield  {author} {\bibinfo {author} {\bibfnamefont {Y.}~\bibnamefont {Liu}}, \bibinfo {author} {\bibfnamefont {Z.}~\bibnamefont {Chen}}, \ and\ \bibinfo {author} {\bibfnamefont {Q.}~\bibnamefont {Shao}},\ }\href {\doibase 10.48550/ARXIV.2412.10888} {\enquote {\bibinfo {title} {Tunable topological states and chirality by non-equilibrium antimagnons in magnetic multilayers},}\ } (\bibinfo {year} {2024})\BibitemShut {NoStop}%
\bibitem [{\citenamefont {Bourhill}\ \emph {et~al.}(2023)\citenamefont {Bourhill}, \citenamefont {Yu}, \citenamefont {Vlaminck}, \citenamefont {Bauer}, \citenamefont {Ruoso},\ and\ \citenamefont {Castel}}]{Bourhill2023}%
  \BibitemOpen
  \bibfield  {author} {\bibinfo {author} {\bibfnamefont {J.}~\bibnamefont {Bourhill}}, \bibinfo {author} {\bibfnamefont {W.}~\bibnamefont {Yu}}, \bibinfo {author} {\bibfnamefont {V.}~\bibnamefont {Vlaminck}}, \bibinfo {author} {\bibfnamefont {G.~E.~W.}\ \bibnamefont {Bauer}}, \bibinfo {author} {\bibfnamefont {G.}~\bibnamefont {Ruoso}}, \ and\ \bibinfo {author} {\bibfnamefont {V.}~\bibnamefont {Castel}},\ }\href {\doibase 10.1103/physrevapplied.19.014030} {\bibfield  {journal} {\bibinfo  {journal} {Physical Review Applied}\ }\textbf {\bibinfo {volume} {19}} (\bibinfo {year} {2023}),\ 10.1103/physrevapplied.19.014030}\BibitemShut {NoStop}%
\bibitem [{\citenamefont {Yu}\ \emph {et~al.}(2020{\natexlab{b}})\citenamefont {Yu}, \citenamefont {Zhang}, \citenamefont {Sharma}, \citenamefont {Blanter},\ and\ \citenamefont {Bauer}}]{TaoYu2020}%
  \BibitemOpen
  \bibfield  {author} {\bibinfo {author} {\bibfnamefont {T.}~\bibnamefont {Yu}}, \bibinfo {author} {\bibfnamefont {X.}~\bibnamefont {Zhang}}, \bibinfo {author} {\bibfnamefont {S.}~\bibnamefont {Sharma}}, \bibinfo {author} {\bibfnamefont {Y.~M.}\ \bibnamefont {Blanter}}, \ and\ \bibinfo {author} {\bibfnamefont {G.~E.~W.}\ \bibnamefont {Bauer}},\ }\href {\doibase 10.1103/physrevb.101.094414} {\bibfield  {journal} {\bibinfo  {journal} {Physical Review B}\ }\textbf {\bibinfo {volume} {101}} (\bibinfo {year} {2020}{\natexlab{b}}),\ 10.1103/physrevb.101.094414}\BibitemShut {NoStop}%
\bibitem [{\citenamefont {Yang}\ \emph {et~al.}(2024)\citenamefont {Yang}, \citenamefont {Ye}, \citenamefont {Su},\ and\ \citenamefont {Wu}}]{Yang:24}%
  \BibitemOpen
  \bibfield  {author} {\bibinfo {author} {\bibfnamefont {T.-L.}\ \bibnamefont {Yang}}, \bibinfo {author} {\bibfnamefont {G.-Z.}\ \bibnamefont {Ye}}, \bibinfo {author} {\bibfnamefont {W.-J.}\ \bibnamefont {Su}}, \ and\ \bibinfo {author} {\bibfnamefont {H.}~\bibnamefont {Wu}},\ }\href {\doibase 10.1364/OL.528451} {\bibfield  {journal} {\bibinfo  {journal} {Opt. Lett.}\ }\textbf {\bibinfo {volume} {49}},\ \bibinfo {pages} {3781} (\bibinfo {year} {2024})}\BibitemShut {NoStop}%
\bibitem [{fer(2005)}]{ferrisphere}%
  \BibitemOpen
  \href {https://www.ferrisphere.com/} {\enquote {\bibinfo {title} {ferrisphere.com},}\ } (\bibinfo {year} {2005})\BibitemShut {NoStop}%
\bibitem [{\citenamefont {Kittel}(1958)}]{Kittel1958}%
  \BibitemOpen
  \bibfield  {author} {\bibinfo {author} {\bibfnamefont {C.}~\bibnamefont {Kittel}},\ }\href {\doibase 10.1103/physrev.112.2139.3} {\bibfield  {journal} {\bibinfo  {journal} {Physical Review}\ }\textbf {\bibinfo {volume} {112}},\ \bibinfo {pages} {2139–2139} (\bibinfo {year} {1958})}\BibitemShut {NoStop}%
\bibitem [{\citenamefont {Nambu}\ \emph {et~al.}(2020)\citenamefont {Nambu}, \citenamefont {Barker}, \citenamefont {Okino}, \citenamefont {Kikkawa}, \citenamefont {Shiomi}, \citenamefont {Enderle}, \citenamefont {Weber}, \citenamefont {Winn}, \citenamefont {Graves-Brook}, \citenamefont {Tranquada}, \citenamefont {Ziman}, \citenamefont {Fujita}, \citenamefont {Bauer}, \citenamefont {Saitoh},\ and\ \citenamefont {Kakurai}}]{Nambu2020}%
  \BibitemOpen
  \bibfield  {author} {\bibinfo {author} {\bibfnamefont {Y.}~\bibnamefont {Nambu}}, \bibinfo {author} {\bibfnamefont {J.}~\bibnamefont {Barker}}, \bibinfo {author} {\bibfnamefont {Y.}~\bibnamefont {Okino}}, \bibinfo {author} {\bibfnamefont {T.}~\bibnamefont {Kikkawa}}, \bibinfo {author} {\bibfnamefont {Y.}~\bibnamefont {Shiomi}}, \bibinfo {author} {\bibfnamefont {M.}~\bibnamefont {Enderle}}, \bibinfo {author} {\bibfnamefont {T.}~\bibnamefont {Weber}}, \bibinfo {author} {\bibfnamefont {B.}~\bibnamefont {Winn}}, \bibinfo {author} {\bibfnamefont {M.}~\bibnamefont {Graves-Brook}}, \bibinfo {author} {\bibfnamefont {J.}~\bibnamefont {Tranquada}}, \bibinfo {author} {\bibfnamefont {T.}~\bibnamefont {Ziman}}, \bibinfo {author} {\bibfnamefont {M.}~\bibnamefont {Fujita}}, \bibinfo {author} {\bibfnamefont {G.}~\bibnamefont {Bauer}}, \bibinfo {author} {\bibfnamefont {E.}~\bibnamefont {Saitoh}}, \ and\ \bibinfo {author} {\bibfnamefont {K.}~\bibnamefont {Kakurai}},\ }\href {\doibase 10.1103/physrevlett.125.027201}
  {\bibfield  {journal} {\bibinfo  {journal} {Physical Review Letters}\ }\textbf {\bibinfo {volume} {125}} (\bibinfo {year} {2020}),\ 10.1103/physrevlett.125.027201}\BibitemShut {NoStop}%
\bibitem [{\citenamefont {Gardin}\ \emph {et~al.}(2023)\citenamefont {Gardin}, \citenamefont {Bourhill}, \citenamefont {Vlaminck}, \citenamefont {Person}, \citenamefont {Fumeaux}, \citenamefont {Castel},\ and\ \citenamefont {Tettamanzi}}]{Gardin2023}%
  \BibitemOpen
  \bibfield  {author} {\bibinfo {author} {\bibfnamefont {A.}~\bibnamefont {Gardin}}, \bibinfo {author} {\bibfnamefont {J.}~\bibnamefont {Bourhill}}, \bibinfo {author} {\bibfnamefont {V.}~\bibnamefont {Vlaminck}}, \bibinfo {author} {\bibfnamefont {C.}~\bibnamefont {Person}}, \bibinfo {author} {\bibfnamefont {C.}~\bibnamefont {Fumeaux}}, \bibinfo {author} {\bibfnamefont {V.}~\bibnamefont {Castel}}, \ and\ \bibinfo {author} {\bibfnamefont {G.~C.}\ \bibnamefont {Tettamanzi}},\ }\href {\doibase 10.1103/physrevapplied.19.054069} {\bibfield  {journal} {\bibinfo  {journal} {Physical Review Applied}\ }\textbf {\bibinfo {volume} {19}} (\bibinfo {year} {2023}),\ 10.1103/physrevapplied.19.054069}\BibitemShut {NoStop}%
\bibitem [{\citenamefont {Gardin}\ \emph {et~al.}(2024)\citenamefont {Gardin}, \citenamefont {Bourcin}, \citenamefont {Bourhill}, \citenamefont {Vlaminck}, \citenamefont {Person}, \citenamefont {Fumeaux}, \citenamefont {Tettamanzi},\ and\ \citenamefont {Castel}}]{Gardin2024}%
  \BibitemOpen
  \bibfield  {author} {\bibinfo {author} {\bibfnamefont {A.}~\bibnamefont {Gardin}}, \bibinfo {author} {\bibfnamefont {G.}~\bibnamefont {Bourcin}}, \bibinfo {author} {\bibfnamefont {J.}~\bibnamefont {Bourhill}}, \bibinfo {author} {\bibfnamefont {V.}~\bibnamefont {Vlaminck}}, \bibinfo {author} {\bibfnamefont {C.}~\bibnamefont {Person}}, \bibinfo {author} {\bibfnamefont {C.}~\bibnamefont {Fumeaux}}, \bibinfo {author} {\bibfnamefont {G.~C.}\ \bibnamefont {Tettamanzi}}, \ and\ \bibinfo {author} {\bibfnamefont {V.}~\bibnamefont {Castel}},\ }\href {\doibase 10.1103/PhysRevApplied.21.064033} {\bibfield  {journal} {\bibinfo  {journal} {Phys. Rev. Appl.}\ }\textbf {\bibinfo {volume} {21}},\ \bibinfo {pages} {064033} (\bibinfo {year} {2024})}\BibitemShut {NoStop}%
\bibitem [{\citenamefont {Resta}(2000)}]{Resta2000}%
  \BibitemOpen
  \bibfield  {author} {\bibinfo {author} {\bibfnamefont {R.}~\bibnamefont {Resta}},\ }\href {\doibase 10.1088/0953-8984/12/9/201} {\bibfield  {journal} {\bibinfo  {journal} {Journal of Physics: Condensed Matter}\ }\textbf {\bibinfo {volume} {12}},\ \bibinfo {pages} {R107–R143} (\bibinfo {year} {2000})}\BibitemShut {NoStop}%
\bibitem [{\citenamefont {Bourcin}\ \emph {et~al.}(2024)\citenamefont {Bourcin}, \citenamefont {Gardin}, \citenamefont {Bourhill}, \citenamefont {Vlaminck},\ and\ \citenamefont {Castel}}]{Bourcin2024}%
  \BibitemOpen
  \bibfield  {author} {\bibinfo {author} {\bibfnamefont {G.}~\bibnamefont {Bourcin}}, \bibinfo {author} {\bibfnamefont {A.}~\bibnamefont {Gardin}}, \bibinfo {author} {\bibfnamefont {J.}~\bibnamefont {Bourhill}}, \bibinfo {author} {\bibfnamefont {V.}~\bibnamefont {Vlaminck}}, \ and\ \bibinfo {author} {\bibfnamefont {V.}~\bibnamefont {Castel}},\ }\href {\doibase 10.1103/physrevapplied.22.064036} {\bibfield  {journal} {\bibinfo  {journal} {Physical Review Applied}\ }\textbf {\bibinfo {volume} {22}} (\bibinfo {year} {2024}),\ 10.1103/physrevapplied.22.064036}\BibitemShut {NoStop}%
\bibitem [{\citenamefont {Walker}(1958)}]{Walker1958}%
  \BibitemOpen
  \bibfield  {author} {\bibinfo {author} {\bibfnamefont {L.~R.}\ \bibnamefont {Walker}},\ }\href {\doibase 10.1063/1.1723117} {\bibfield  {journal} {\bibinfo  {journal} {Journal of Applied Physics}\ }\textbf {\bibinfo {volume} {29}},\ \bibinfo {pages} {318–323} (\bibinfo {year} {1958})}\BibitemShut {NoStop}%
\bibitem [{\citenamefont {Harder}\ \emph {et~al.}(2016)\citenamefont {Harder}, \citenamefont {Hyde}, \citenamefont {Bai}, \citenamefont {Match},\ and\ \citenamefont {Hu}}]{Harder2016}%
  \BibitemOpen
  \bibfield  {author} {\bibinfo {author} {\bibfnamefont {M.}~\bibnamefont {Harder}}, \bibinfo {author} {\bibfnamefont {P.}~\bibnamefont {Hyde}}, \bibinfo {author} {\bibfnamefont {L.}~\bibnamefont {Bai}}, \bibinfo {author} {\bibfnamefont {C.}~\bibnamefont {Match}}, \ and\ \bibinfo {author} {\bibfnamefont {C.-M.}\ \bibnamefont {Hu}},\ }\href {\doibase 10.1103/physrevb.94.054403} {\bibfield  {journal} {\bibinfo  {journal} {Physical Review B}\ }\textbf {\bibinfo {volume} {94}} (\bibinfo {year} {2016}),\ 10.1103/physrevb.94.054403}\BibitemShut {NoStop}%
\bibitem [{\citenamefont {Zhang}\ \emph {et~al.}(2015)\citenamefont {Zhang}, \citenamefont {Zou}, \citenamefont {Zhu}, \citenamefont {Marquardt}, \citenamefont {Jiang},\ and\ \citenamefont {Tang}}]{Zhang2015}%
  \BibitemOpen
  \bibfield  {author} {\bibinfo {author} {\bibfnamefont {X.}~\bibnamefont {Zhang}}, \bibinfo {author} {\bibfnamefont {C.-L.}\ \bibnamefont {Zou}}, \bibinfo {author} {\bibfnamefont {N.}~\bibnamefont {Zhu}}, \bibinfo {author} {\bibfnamefont {F.}~\bibnamefont {Marquardt}}, \bibinfo {author} {\bibfnamefont {L.}~\bibnamefont {Jiang}}, \ and\ \bibinfo {author} {\bibfnamefont {H.~X.}\ \bibnamefont {Tang}},\ }\href {\doibase 10.1038/ncomms9914} {\bibfield  {journal} {\bibinfo  {journal} {Nature Communications}\ }\textbf {\bibinfo {volume} {6}} (\bibinfo {year} {2015}),\ 10.1038/ncomms9914}\BibitemShut {NoStop}%
\bibitem [{\citenamefont {Fan}\ \emph {et~al.}(2024)\citenamefont {Fan}, \citenamefont {Zuo}, \citenamefont {Li},\ and\ \citenamefont {Li}}]{https://doi.org/10.48550/arxiv.2401.02280}%
  \BibitemOpen
  \bibfield  {author} {\bibinfo {author} {\bibfnamefont {Z.-Y.}\ \bibnamefont {Fan}}, \bibinfo {author} {\bibfnamefont {X.}~\bibnamefont {Zuo}}, \bibinfo {author} {\bibfnamefont {H.-T.}\ \bibnamefont {Li}}, \ and\ \bibinfo {author} {\bibfnamefont {J.}~\bibnamefont {Li}},\ }\href {\doibase 10.48550/ARXIV.2401.02280} {\enquote {\bibinfo {title} {Nonreciprocal entanglement in cavity magnomechanics exploiting chiral cavity-magnon coupling},}\ } (\bibinfo {year} {2024})\BibitemShut {NoStop}%
\bibitem [{\citenamefont {Aiello}\ \emph {et~al.}(2015)\citenamefont {Aiello}, \citenamefont {Banzer}, \citenamefont {Neugebauer},\ and\ \citenamefont {Leuchs}}]{Aiello2015}%
  \BibitemOpen
  \bibfield  {author} {\bibinfo {author} {\bibfnamefont {A.}~\bibnamefont {Aiello}}, \bibinfo {author} {\bibfnamefont {P.}~\bibnamefont {Banzer}}, \bibinfo {author} {\bibfnamefont {M.}~\bibnamefont {Neugebauer}}, \ and\ \bibinfo {author} {\bibfnamefont {G.}~\bibnamefont {Leuchs}},\ }\href {\doibase 10.1038/nphoton.2015.203} {\bibfield  {journal} {\bibinfo  {journal} {Nature Photonics}\ }\textbf {\bibinfo {volume} {9}},\ \bibinfo {pages} {789–795} (\bibinfo {year} {2015})}\BibitemShut {NoStop}%
\bibitem [{\citenamefont {Aiello}\ and\ \citenamefont {Banzer}(2016)}]{Aiello2016}%
  \BibitemOpen
  \bibfield  {author} {\bibinfo {author} {\bibfnamefont {A.}~\bibnamefont {Aiello}}\ and\ \bibinfo {author} {\bibfnamefont {P.}~\bibnamefont {Banzer}},\ }\href {\doibase 10.1088/2040-8978/18/8/085605} {\bibfield  {journal} {\bibinfo  {journal} {Journal of Optics}\ }\textbf {\bibinfo {volume} {18}},\ \bibinfo {pages} {085605} (\bibinfo {year} {2016})}\BibitemShut {NoStop}%
\bibitem [{\citenamefont {Bliokh}\ and\ \citenamefont {Nori}(2015)}]{Bliokh2015}%
  \BibitemOpen
  \bibfield  {author} {\bibinfo {author} {\bibfnamefont {K.~Y.}\ \bibnamefont {Bliokh}}\ and\ \bibinfo {author} {\bibfnamefont {F.}~\bibnamefont {Nori}},\ }\href {\doibase 10.1016/j.physrep.2015.06.003} {\bibfield  {journal} {\bibinfo  {journal} {Physics Reports}\ }\textbf {\bibinfo {volume} {592}},\ \bibinfo {pages} {1–38} (\bibinfo {year} {2015})}\BibitemShut {NoStop}%
\bibitem [{\citenamefont {Bai}\ \emph {et~al.}(2015)\citenamefont {Bai}, \citenamefont {Harder}, \citenamefont {Chen}, \citenamefont {Fan}, \citenamefont {Xiao},\ and\ \citenamefont {Hu}}]{Bai2015}%
  \BibitemOpen
  \bibfield  {author} {\bibinfo {author} {\bibfnamefont {L.}~\bibnamefont {Bai}}, \bibinfo {author} {\bibfnamefont {M.}~\bibnamefont {Harder}}, \bibinfo {author} {\bibfnamefont {Y.}~\bibnamefont {Chen}}, \bibinfo {author} {\bibfnamefont {X.}~\bibnamefont {Fan}}, \bibinfo {author} {\bibfnamefont {J.}~\bibnamefont {Xiao}}, \ and\ \bibinfo {author} {\bibfnamefont {C.-M.}\ \bibnamefont {Hu}},\ }\href {\doibase 10.1103/physrevlett.114.227201} {\bibfield  {journal} {\bibinfo  {journal} {Physical Review Letters}\ }\textbf {\bibinfo {volume} {114}} (\bibinfo {year} {2015}),\ 10.1103/physrevlett.114.227201}\BibitemShut {NoStop}%
\end{thebibliography}%

\appendix

\onecolumngrid

\section{System's design and characterization}\phantomsection\label{appendixA}

\subsection{Cavity's design and fabrication}\label{appendixA1}

Our cavity is based on substrate integrated waveguide (SIW) technology, which is often used in integrated microwave devices. Its main feature is its lateral confinement of the electromagnetic field by means of metal-plated arrays of vias. If the vias' spacing is smaller than half the wavelength, then the arrays will act as walls for the cavity photons. By using a dielectric instead of air like usual three-dimensional cavities, we reduce the cavity volume significantly allowing to enhance the magnon-photon coupling. We showed that the resonant frequency of the cavity is $11$ GHz which would correspond to a $10$ mm wavelength, knowing that the spacing of the vias is $420$ $\mu m$ with a vias radius of $250$ $\mu m$, we ensure a good confinement of the cavity photons.

Our device is constituted of an upper and lower metalized Rogers\textsuperscript{®} RO3003 $\epsilon_r = 3$ substrate of $h=1.52$ mm thickness. The dielectric is metalized with pure laminated copper with a thickness of $35$ $\mu m$. The device is then drilled before its metallization to create the vias, adding approximately $15$ $\mu m$ to the metallic surface. The upper plane is then etched to create the coupling loops at the input and output of the resonator.

\begin{figure}[hbt!]
    \centering
    \includegraphics[width=0.6\linewidth]{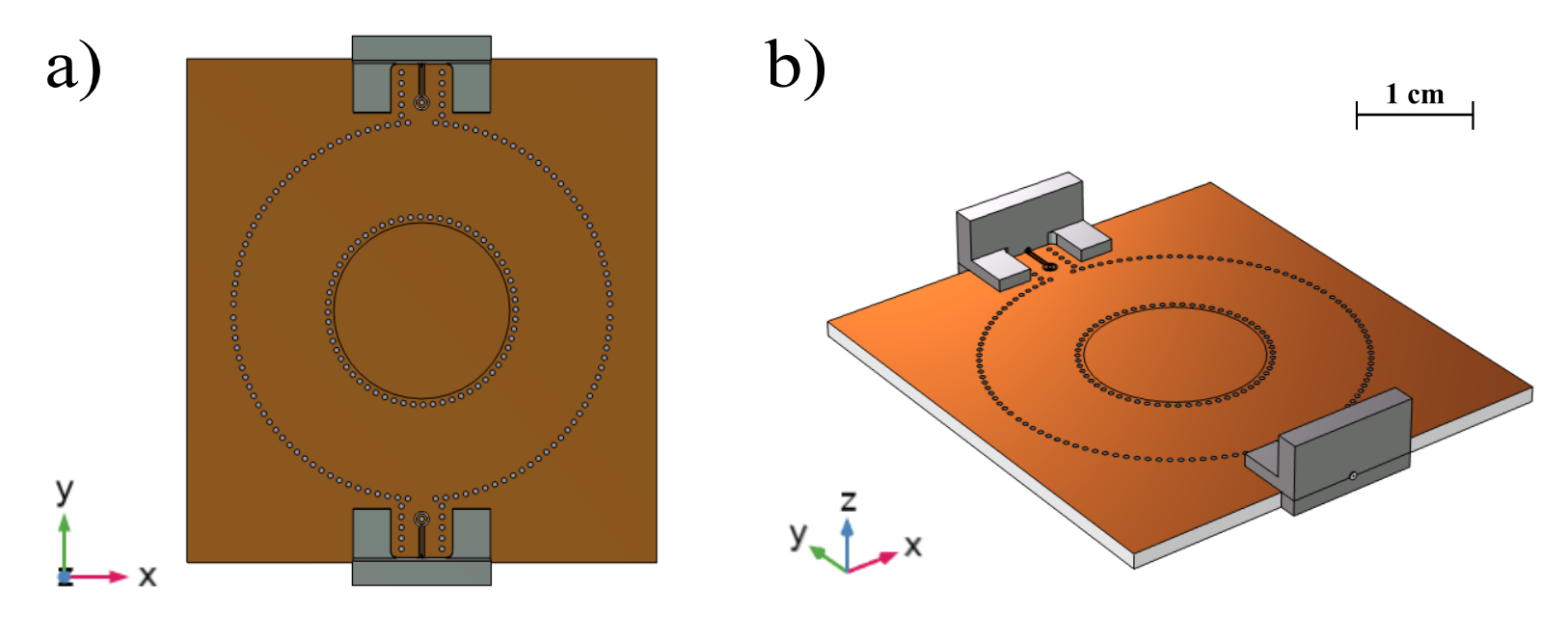}
    \caption{\textbf{a)} top and \textbf{b)} side view of the SIW chiral resonator.}
    \label{APP_FIG_5}
\end{figure}

We use a vector network analyzer (VNA) to characterize the device as shown in Figure \ref{FIG_1} \textbf{a)}. We measure all 4 components of the S-matrix: $S_{11}$, $S_{12}$, $S_{21}$ and $S_{22}$ allowing us to easily determine the isolation ratio quantifying the non-reciprocity of the system. The signal coming from the VNA is coupled to the chiral cavity by means of two coupling loops etched from the top and bottom metallic surface of the SIW. Two Southwest\textsuperscript{®} connectors, with coaxial ends are coupled to the loops. Several holes located at different chiral and non-chiral positions are drilled to host the $1$ mm YIG spheres which are collectively brought to resonance using a bias magnetic field applied along the $z$-axis.

\subsection{FEM modeling}\label{appendixA2}

According to the Maxwell equations, the two polar components of the magnetic field inside the cavity $\textbf{H}=H_{\rho}\textbf{e}_{\rho}+H_{\phi}\textbf{e}_{\phi}$ are written:

\begin{align}
    H_{\rho}(\rho, \phi) &=\frac{1}{\mu_0\gamma c} \frac{m}{\rho} E_z,\\
    H_{\phi}(\rho, \phi) &=-i \frac{1}{\mu_0\gamma c} \frac{\partial E_z}{\partial \rho}.
\end{align}

At specific azimuthal lines of the resonator, the magnetic field of the TE modes is circularly polarized, verifying the relations $H\textsubscript{$\phi$}=iH\textsubscript{$\rho$}$ for right-hand polarization (position $\rho\textsubscript{+}$, $m>0$) and $H\textsubscript{$\phi$}=-iH\textsubscript{$\rho$}$ (position $\rho\textsubscript{-}$, $m<0$) for left-hand polarization. Position $\rho\textsubscript{0}$ is the special position for which the linearly polarized $E\textsubscript{z}$ field is maximum, verifying the relation $H\textsubscript{$\phi$}(\rho\textsubscript{0},\phi)=0$ (the special positions $\rho\textsubscript{0}$ and $\rho\textsubscript{$\pm$}$ are displayed in Figure \ref{FIG_1} \textbf{b)}).

\begin{figure}[hbt!]
    \centering
    \includegraphics[width=\linewidth]{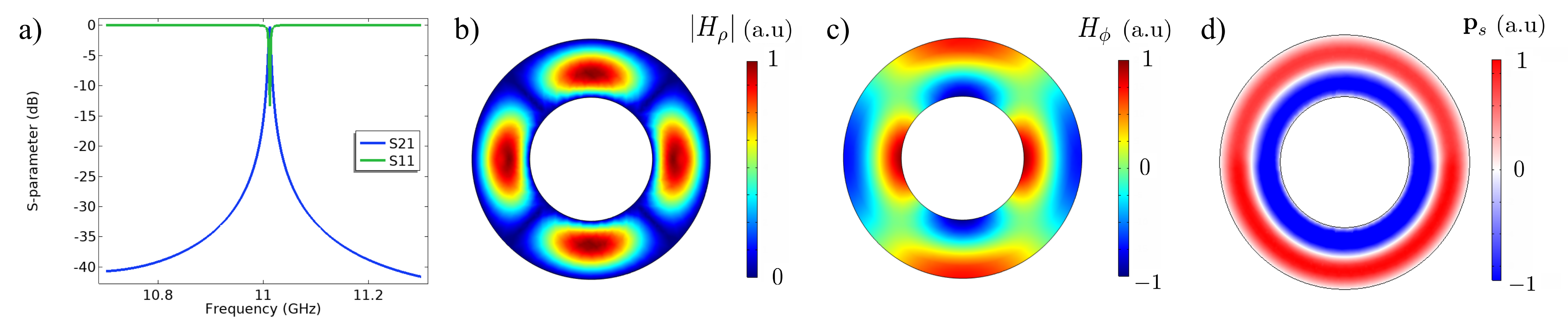}
    \caption{COMSOL\textsuperscript{®} modeling of the chiral cavity. \textbf{a)} Transmission ($S_{21}$ in blue) and reflection ($S_{11}$ in green) parameter of the empty cavity in dB. \textbf{b)}-\textbf{d} Distribution of $H_{\rho}$, $H_{\phi}$ and $\textbf{p}_s$ inside the resonator in the $(\rho,\phi)$ plane at $z=h/2$ respectively.}
    \label{APP_FIG_6}
\end{figure}

Our simulation results were derived using the FEM software COMSOL Multiphysics\textsuperscript{®}. The Maxwell equations, were solved with the electromagnetic waves (emw) module, allowing us to characterize the distribution of the electric and magnetic field along with the reflection and transmission parameter of the empty cavity's resonant mode (c.f. Figure \ref{APP_FIG_6}). The results of this Appendix were obtained through frequency domain and eigenfrequency studies. The geometry used in COMSOL\textsuperscript{®} to model our cavity is not reproducing the substrate integrated waveguide technology of our device. Instead of holes, we define walls to confine the electromagnetic signal. This simplification is due to the meshing complexity of such vast arrays of sub-millimeter holes, allowing to significantly reduce computing time. Still, we find good consistency between our measurements and simulation.

One can also compute the spin angular momentum of the cavity's photons $\textbf{p}_s$ to highlight the chiral properties of the cavity's geometry. The spin angular momentum is the spin contribution of the linear momentum density written: $\textbf{p}=\textbf{p}_0+\textbf{p}_s$, where $\textbf{p}_0$ is the orbital contribution \cite{Bourhill2023}. The spin angular momentum reads

\begin{equation}
    \textbf{p}_s = \frac{1}{8\omega}\nabla\times \mathfrak{Im}[\epsilon_0\textbf{E}^*\times\textbf{E}+\mu_0\textbf{H}^*\times\textbf{H}]
\end{equation}

with $\epsilon_0$ and $\mu_0$ the vacuum permittivity and permeability respectively. It is defined as the local degree of microwave circular polarization which has a node at $\rho_0$, becomes positive at $\rho>\rho_0$ and negative at $\rho<\rho_0$ as shown in Figure \ref{APP_FIG_12} \textbf{d)}. In our system the spin angular momentum is perpendicular to the signal propagation, this transverse orientation of the photon's spin is also known as "photonic wheel" in optics \cite{Aiello2015,Aiello2016,Bliokh2015,Bliokh2017}.

\begin{figure}[hbt!]
    \centering
    \includegraphics[width=0.7\linewidth]{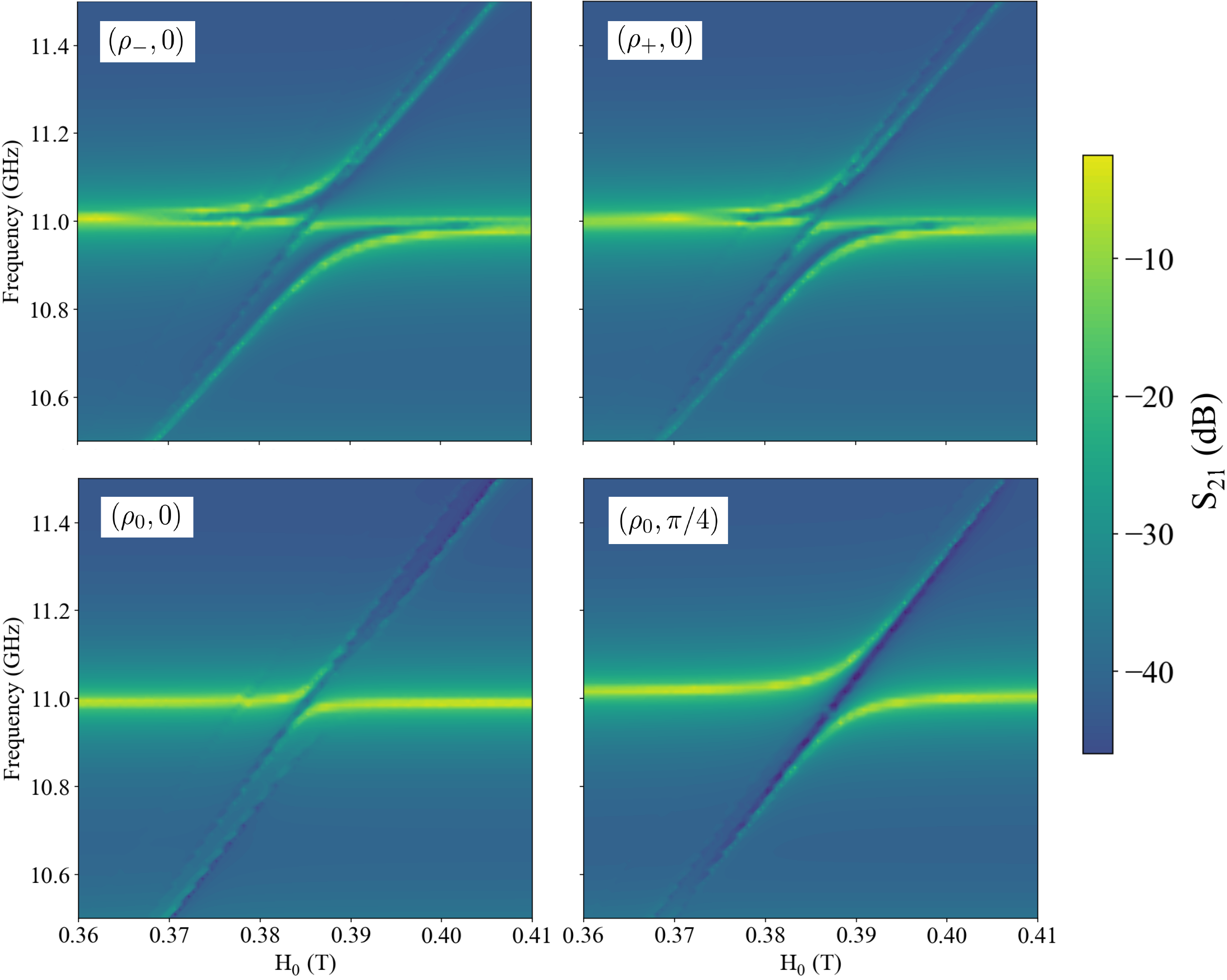}
    \caption{\textbf{a)}-\textbf{d)} Transmission spectrum $S_{21}$ calculated numerically for a single YIG sphere loaded in the cavity at different positions (insets).}
    \label{APP_FIG_7}
\end{figure}

In order to compute the spin photon interaction on COMSOL\textsuperscript{®}, we modeled the magnetic part using a relative permeability tensor, also known as Polder tensor, considering the Gilbert damping, saturation magnetization and complex permittivity. By solving the linearized Landau-Lifshitz-Gilbert (LLG) equation coupled with the Maxwell equations and by making the rotating-wave and macrospin approximations, one can write the Polder tensor as:

\begin{equation}
\mu\textsubscript{r}=
   \begin{pmatrix}
    1+\chi_{xx} & \chi_{xy} & 0 \\
    -\chi_{yx} & 1+\chi_{yy} & 0\\
    0 & 0 & 1
\end{pmatrix},
\end{equation}

with:

\begin{align}
    \chi_{xx} &= \frac{\omega_0\omega_M(\omega_0^2-\omega^2)+\omega_0\omega_M\omega^2\alpha^2}{(\omega_0^2-\omega^2(1+\alpha^2))^2+(2\omega_0\omega\alpha)^2} - i\frac{\alpha\omega\omega_M(\omega_0^2-\omega^2(1+\alpha^2))}{(\omega_0^2-\omega^2(1+\alpha^2))^2+(2\omega_0\omega\alpha)^2}\\
    \chi_{xy} &= \frac{2\omega_0\omega_M\omega^2\alpha}{(\omega_0^2-\omega^2(1+\alpha^2))^2+(2\omega_0\omega\alpha)^2} + i\frac{\omega\omega_M(\omega_0^2-\omega^2(1+\alpha^2))}{(\omega_0^2-\omega^2(1+\alpha^2))^2+(2\omega_0\omega\alpha)^2}\\
\end{align}

and where $\chi_{yy} = \chi_{xx}$, $\chi_{yx} = \chi_{xy}$, $\omega_M=\gamma M_s$ and $\omega_0=\gamma (H_0-M_s/3)$. This frequency and field-dependent tensor is used to compute the coupling of the magnetization to the cavity mode. Using this model, we are able to determine transmission and reflection parameters according to frequency and applied magnetic field, allowing to determine the coupling strength of the CMP (see Figure \ref{APP_FIG_7}).

\section{Derivation of theory}\label{appendixB}

\subsection{Calculation of the interaction Hamiltonian}\label{appendixB1}

In our chiral cavity, the magnetic field of the TE mode is quantized by the photon operator $\hat{\alpha}_m$ and can be written in this manner:

\begin{equation}
    \hat{\textbf{H}}(\textbf{r}) = \sum_m[\mathcal{H}^m(\rho,z)e^{im\phi}\hat{\alpha}_m + (\mathcal{H}^m(\rho,z))^*e^{-im\phi}\hat{\alpha}_m^{\dag}],
\end{equation}

where $\mathcal{H}_m(\rho,z)$ is the normalized magnetic field that can be written in terms of polar components,

\begin{equation}
    \mathcal{H}^m(\rho,z)=\mathcal{H}_{\rho}^m(\rho,z)\hat{\textbf{e}}_{\rho} + \mathcal{H}_{\phi}^m(\rho,z)\hat{\textbf{e}}_{\phi},
\end{equation}

with $\mathcal{H}^m(\rho,\phi,z)=\frac{\textbf{H}(\rho,z)}{\mathcal{N}}$ where $\mathcal{N}$ is a normalization constant and $\textbf{H}(\rho,\phi,z)$ is the magnetic field which is written in the polar basis as: 

\begin{equation}
    \textbf{H}(\rho,\phi,z)=H_{\rho}(\rho,\phi,z)\hat{\textbf{e}}_{\rho} + H_{\phi}(\rho,\phi,z)\hat{\textbf{e}}_{\phi},
    \label{mag_field}
\end{equation}

according to the Maxwell equations the two components read

\begin{align}
H\textsubscript{$\rho$}(\rho,\phi, z) &=  \frac{1}{\mu\textsubscript{0}\gamma\textsubscript{m}c} \frac{m}{\rho} E\textsubscript{z}, \\ 
H\textsubscript{$\phi$}(\rho,\phi, z) &=  -i \frac{1}{\mu\textsubscript{0}\gamma\textsubscript{m}c} \frac{m}{\rho} \frac{\partial E\textsubscript{z}}{\partial\rho}.
\end{align}

By considering the clockwise $(\phi)$ or counterclockwise $(-\phi)$ propagation of the signal inside the cavity, equation (\ref{mag_field}) rewrites:

\begin{equation}
    \textbf{H}(\rho,s\phi,z)=[H_{\rho}(\rho,z)\hat{\textbf{e}}_{\rho} + H_{s\phi}(\rho,z)\hat{\textbf{e}}_{s\phi}]e^{im\phi} = [H_{\rho}(\rho,z)\hat{\textbf{e}}_{\rho} + sH_{\phi}(\rho,z)\hat{\textbf{e}}_{\phi}]e^{im\phi}
    \label{full_mag_field}
\end{equation}

with $s$=sgn$(m)$ and where the properties $\hat{\textbf{e}}_{-\phi}=-\hat{\textbf{e}}_{\phi}$ and $H_{-\phi}(\rho,z)=H_{\phi}(\rho,z)$ were used. In order to determine the expression of the Zeeman interaction, we use the following expression for the spin operator: $\hat{\textbf{S}}_{\perp} = \hat{S}_x\hat{\textbf{e}}_x + \hat{S}_y\hat{\textbf{e}}_y$, where 

\begin{equation}
    \hat{S}_x = \frac{\hbar}{2}\sqrt{2S}(\hat{m}^{\dag}+\hat{m}), \hspace{0.3cm} \hat{S}_y = i\frac{\hbar}{2}\sqrt{2S}(\hat{m}^{\dag}-\hat{m}).
\end{equation}

The Zeeman interaction is written

\begin{equation}
    \hat{H}_{int} = -\mu_0\sum_{l=1}^{N}\int d\textbf{r}\hat{\textbf{H}}\cdot\hat{\textbf{M}}_{l,s},
\end{equation}

where: $\hat{\textbf{M}}_{l,s} = -\gamma \hbar \hat{\textbf{S}}_{l}$. After a projection in the Cartesian basis of expression \ref{full_mag_field} and under rotating wave approximation, the Zeeman interaction finally reads:

\begin{equation}
    \hat{H}_{int} = \mu_0\hbar\sqrt{\frac{\gamma M_{l,s}V_l}{2\hbar}}\sum_m\sum_{l=1}^N(\mathcal{H}_{\rho}^{m}+is\mathcal{H}_{\phi}^m)e^{i(sm_*\phi+\theta_m)}\hat{m}_l^{\dag}\hat{\alpha}_m + H.c
\end{equation}

where we can identify the coupling strength: $g_{l,m}=\mu_0 \sqrt{\frac{\gamma M_{l,s} V_l}{2 \hbar}}(\mathcal{H}^m_{\rho}+is\mathcal{H}^m_{\phi})$ and coupling phase: $\theta_m=\atantwo(s \frac{\mathcal{H}^m_{\phi}}{\mathcal{H}^m_{\rho}})$

\subsection{Input-output model for a single YIG sphere}\label{appendixB2}

We recall the expression of the system's free Hamiltonian:

\begin{equation}
    \frac{\hat{H}_{free}}{\hbar}=\Tilde{\omega}_+\hat{\alpha}_{+}^\dag\hat{\alpha}_{+}+\Tilde{\omega}_-\hat{\alpha}_{-}^\dag\hat{\alpha}_{-}+\Tilde{\omega}_{K}\hat{m}^\dag\hat{m},
    \label{H_free}
\end{equation}

along with the interaction Hamiltonian:

\begin{equation}
    \frac{\hat{H}_{int}}{\hbar}=G_+\hat{\alpha}_{+}\hat{m}^\dag + G_-\hat{\alpha}_{-}\hat{m}^\dag + H.c.,
    \label{H_int}
\end{equation}

where ${G}_{\pm}=|g_{\pm}(\rho)|e^{\mp 3i\phi}e^{i\theta_s}$ corresponds to the complex coupling strength with $g_{\pm}$ the expression of the coupling strength with CW $(m=2, g_+)$ or CCW cavity mode $(m=-2, g_-)$ simplified for notations. $\Tilde{\omega}_{\pm}=\omega_{\pm}-i\kappa_c/2$ and $\Tilde{\omega}_K=\omega_K-i\kappa_m/2$, with $\kappa_c$ and $\kappa_m$ are the intrinsic loss rates of the cavity modes and the magnon mode respectively. For the magnon, the intrinsic loss rate reads: $\kappa_m=\alpha\omega_K$, with $\alpha$ the Gilbert damping rate. The sum of equations (\ref{H_free}) and (\ref{H_int}) gives the Hamiltonian of the system, now we need to add the contribution of the ports. This can be done by considering the so-called "bath" operators introduced by input-output formalism. To each port is attributed two bath operators: $\hat{b}_{1}^{in}$, $\hat{b}_{1}^{out}$ for port one and $\hat{b}_{2}^{in}$, $\hat{b}_{2}^{out}$ for port 2. Since bath operators only couple to cavity modes, we will find their contribution only in the equation of motion of the photonic modes operators:

\begin{align}
    \frac{d \hat{\alpha}_-}{dt} &=-i\Tilde{\omega}_-\hat{\alpha}_- - iG_-^*\hat{m} - \kappa_{ccw,1}^*\left(\hat{b}_1^{in} + \frac{1}{2}\kappa_{cw,1}\hat{\alpha}_+ + \frac{1}{2}\kappa_{ccw,1}\hat{\alpha}_-\right) - \kappa_{ccw,2}^*\left(\hat{b}_2^{in}+\frac{1}{2}\kappa_{cw,2}\hat{\alpha}_+ + \frac{1}{2}\kappa_{ccw,2}\hat{\alpha}_-\right),\\
   \frac{d \hat{\alpha}_+}{dt} &=-i\Tilde{\omega}_+\hat{\alpha}_+ - iG_+^*\hat{m} - \kappa_{cw,1}^*\left(\hat{b}_1^{in} + \frac{1}{2}\kappa_{cw,1}\hat{\alpha}_+ + \frac{1}{2}\kappa_{ccw,1}\hat{\alpha}_-\right) - \kappa_{cw,2}^*\left(\hat{b}_2^{in}+\frac{1}{2}\kappa_{cw,2}\hat{\alpha}_+ + \frac{1}{2}\kappa_{ccw,2}\hat{\alpha}_-\right), \\
   \frac{d \hat{m}}{dt} &=-i\Tilde{\omega}_K\hat{m} - iG_-\hat{\alpha}_- - iG_+\hat{\alpha}_+.
\end{align}

After taking the Fourier transform of the set of equations, it reads:

\begin{align}
    \label{alpha_-}
    \omega\hat{\alpha}_- &=\Tilde{\omega}_-\hat{\alpha}_- + G_-^*\hat{m}- i\kappa_{ccw,1}^*\left(\hat{b}_1^{in}+\frac{1}{2}\kappa_{cw,1}\hat{\alpha}_+ + \frac{1}{2}\kappa_{ccw,1}\hat{\alpha}_- \right) - i\kappa_{ccw,2}^*\left(\hat{b}_2^{in} + \frac{1}{2}\kappa_{cw,2}\hat{\alpha}_+ + \frac{1}{2}\kappa_{ccw,2}\hat{\alpha}_-\right)\\
    \label{alpha_+}
    \omega\hat{\alpha}_+ &=\Tilde{\omega}_+\hat{\alpha}_+ + G_+^*\hat{m}- i\kappa_{cw,1}^*\left(\hat{b}_1^{in}+\frac{1}{2}\kappa_{cw,1}\hat{\alpha}_+ + \frac{1}{2}\kappa_{ccw,1}\hat{\alpha}_- \right) - i\kappa_{cw,2}^*\left(\hat{b}_2^{in} + \frac{1}{2}\kappa_{cw,2}\hat{\alpha}_+ + \frac{1}{2}\kappa_{ccw,2}\hat{\alpha}_-\right)\\
    \omega \hat{m} &=\Tilde{\omega}_K\hat{m} + G_-\hat{\alpha}_- + iG_+\hat{\alpha}_+.
    \label{omega_m_dynamic}
\end{align}

\begin{figure}
    \centering
    \includegraphics[width=0.6\linewidth]{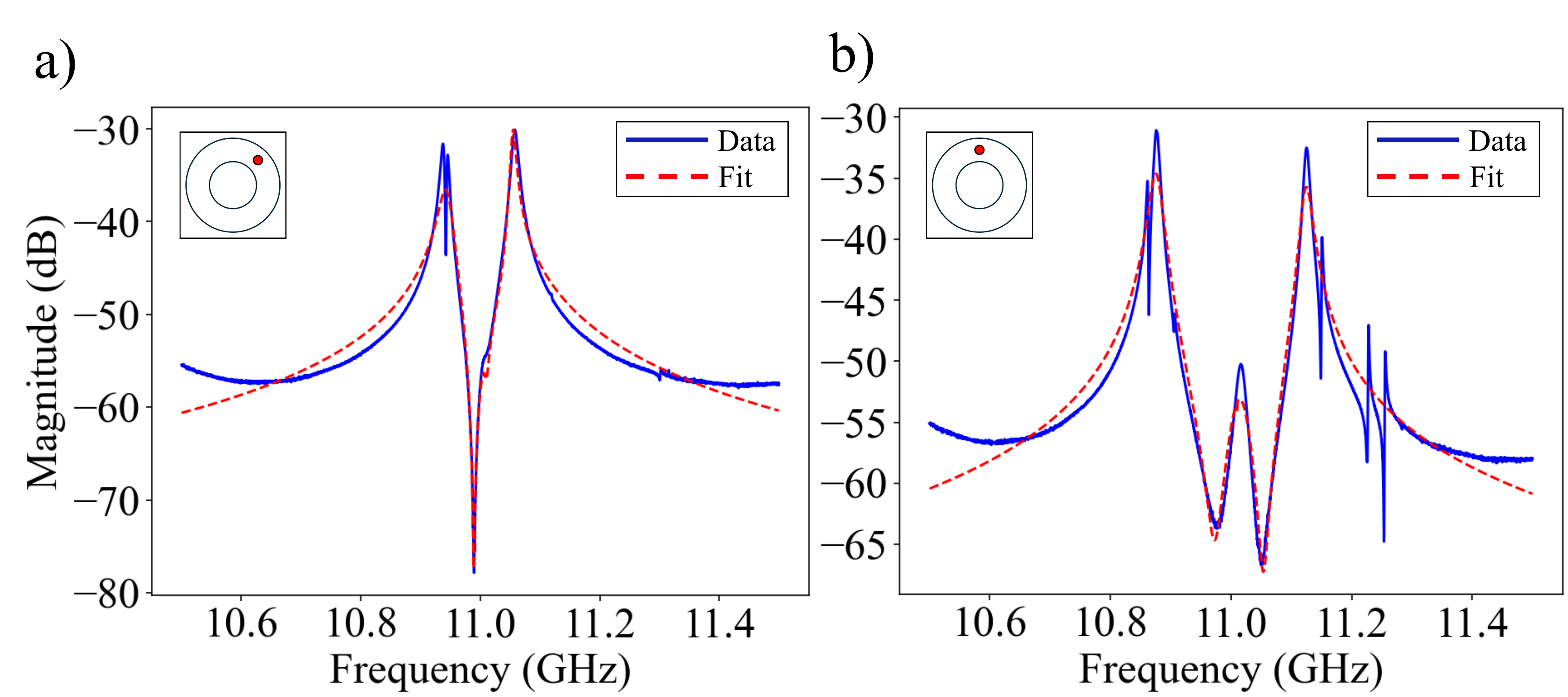}
    \caption{\textbf{a)}, \textbf{b)} Measured (blue) $S_{21}$ transmission parameter for a single YIG sphere loaded cavity located at \textbf{a)} $(\rho_0, \pi/4)$ and \textbf{b)} $(\rho=10.2 \text{mm}, 0)$ at resonance (see insets). The sharp dip in magnitude corresponds to the anti-resonance. The dashed red lines correspond to the fitting using equation (\ref{S21_appendix}).}
    \label{APP_FIG_8}
\end{figure}

Equation (\ref{omega_m_dynamic}) gives:

\begin{equation}
    \hat{m}=\frac{G_+}{\omega-\Tilde{\omega}_K}\hat{\alpha}_+ + \frac{G_-}{\omega-\Tilde{\omega}_K}\hat{\alpha}_-,
\end{equation}

then, we inject this expression in equation (\ref{alpha_-}):

\begin{equation}
    \left(\omega - \Tilde{\omega}_-^{'} - \frac{|G_-|^2}{\omega-\Tilde{\omega}_K}\right)\hat{\alpha}_- = 
    \left(\frac{G_+ G_-^*}{\omega - \Tilde{\omega}_K}-\frac{i}{2}J\right)\hat{\alpha}_+ 
    - i\left(\kappa_{ccw,1}^*\hat{b}_1^{in}+\kappa_{ccw,2}^*\hat{b}_2^{in}\right),
\end{equation}

where: $\Tilde{\omega}_-^{'}=\Tilde{\omega}_- - \frac{i}{2}(|\kappa_{ccw,1}|^2 + |\kappa_{ccw,2}|^2)$ and $J=(\kappa_{ccw,1}^*\kappa_{cw,1} + \kappa_{ccw,2}^*\kappa_{cw,2})$. The latter corresponds to an indirect coupling between the two photonic probes that has been artificially added in previous models \cite{Bourhill2023} but never explicitly defined. Similarly, by injecting equation (\ref{omega_m_dynamic}) in (\ref{alpha_+}) we obtain:

\begin{equation}
    \left(\omega - \Tilde{\omega}_+^{'} - \frac{|G_+|^2}{\omega-\Tilde{\omega}_K}\right)\hat{\alpha}_+ = 
    \left(\frac{G_+^* G_-}{\omega - \Tilde{\omega}_K}-\frac{i}{2}J^*\right)\hat{\alpha}_- \\
    - i\left(\kappa_{cw,1}^*\hat{b}_1^{in}+\kappa_{cw,2}^*\hat{b}_2^{in}\right),
\end{equation}

where: $\Tilde{\omega}_+^{'}=\Tilde{\omega}_+ - \frac{i}{2}(|\kappa_{cw,1}|^2 + |\kappa_{cw,2}|^2)$. Now that we have defined the two coupled equations for $\hat{\alpha}_-$ and $\hat{\alpha}_+$, we can inject one in the other to find: 

\begin{align}
    \hat{\alpha}_- &= -i \frac{\chi_{ccw}^*\left(\frac{G_+^*G_-}{\omega - \Tilde{\omega}_K}-\frac{i}{2}J^*\right) + \chi_{cw}^*\Gamma_-}{\Gamma_+\Gamma_- - \left(\frac{G_+G_-^*}{\omega - \Tilde{\omega}_K} - \frac{i}{2}J\right) \left(\frac{G_+^*G_-}{\omega - \Tilde{\omega}_K} - \frac{i}{2}J^*\right)},\\
    \hat{\alpha}_+ &= -i \frac{\chi_{cw}^*\left(\frac{G_+G_-^*}{\omega - \Tilde{\omega}_K}-\frac{i}{2}J\right) + \chi_{ccw}^*\Gamma_+}{\Gamma_+\Gamma_- - \left(\frac{G_+G_-^*}{\omega - \Tilde{\omega}_K} - \frac{i}{2}J\right) \left(\frac{G_+^*G_-}{\omega - \Tilde{\omega}_K} - \frac{i}{2}J^*\right)},
\end{align}

with $\chi_{cw}=\kappa_{cw,1}\hat{b}_1^{in} + \kappa_{cw,2}\hat{b}_2^{in}$, $\chi_{ccw}=\kappa_{ccw,1}\hat{b}_1^{in} + \kappa_{ccw,2}\hat{b}_2^{in}$ and $\Gamma_{\pm} = \omega - \Tilde{\omega}_{\pm}^{'} - \frac{|G_{\pm}|^2}{\omega - \Tilde{\omega}_{\pm}^{'}}$. At this point, we are ready to calculate the transmission parameters $S_{21}$ and $S_{12}$ defined as:

\begin{equation}
    S_{21}=\left.\frac{\hat{b}_2^{out}}{\hat{b}_1^{in}}\right|_{\hat{b}_2^{in}=0}, \hspace{0.3cm} S_{12}=\left.\frac{\hat{b}_1^{out}}{\hat{b}_2^{in}}\right|_{\hat{b}_1^{in}=0},
\end{equation}

knowing the input-output relations of our system:

\begin{align}
    \hat{b}_1^{out} &= \hat{b}_1^{in} + \kappa_{cw,1}\hat{\alpha}_+ + \kappa_{ccw,1}\hat{\alpha}_-,\\
    \hat{b}_2^{out} &= \hat{b}_2^{in} + \kappa_{cw,2}\hat{\alpha}_+ + \kappa_{ccw,2}\hat{\alpha}_-,
\end{align}

we finally obtain the full expression of the two transmission parameters:

\begin{equation}
    S_{21} = -i \frac{\kappa_{cw2}[\kappa_{ccw1}^*(\frac{G_+^*G_-}{\omega - \omega_K}-\frac{i}{2}J^*) + \kappa_{cw1}^*\Gamma_-] + \kappa_{ccw2}[\kappa_{cw1}^*(\frac{G_+G_-^*}{\omega - \omega_K}-\frac{i}{2}J) + \kappa_{ccw1}^*\Gamma_+]}{\Gamma_-\Gamma_+ - (\frac{G_+G_-^*}{\omega - \omega_K}-\frac{i}{2}J) (\frac{G_+^*G_-}{\omega - \omega_K}-\frac{i}{2}J^*)},
    \label{S21_appendix}
\end{equation}

and:

\begin{equation}
    S_{12} = -i \frac{\kappa_{cw1}[\kappa_{ccw2}^*(\frac{G_+^*G_-}{\omega - \omega_K}-\frac{i}{2}J^*) + \kappa_{cw2}^*\Gamma_-] + \kappa_{ccw1}[\kappa_{cw2}^*(\frac{G_+G_-^*}{\omega - \omega_K}-\frac{i}{2}J) + \kappa_{ccw2}^*\Gamma_+]}{\Gamma_-\Gamma_+ - (\frac{G_+G_-^*}{\omega - \omega_K}-\frac{i}{2}J) (\frac{G_+^*G_-}{\omega - \omega_K}-\frac{i}{2}J^*)},
\end{equation}

based on those expressions of the transmission amplitude, we can achieve a fitting of our experimental data and evidence the shape of the anti-resonances observed in \ref{FIG_3} \textbf{a)}-\textbf{c)}. In the case where the YIG sphere is located at $(\rho_0, \pi/4)$ both CW and CCW cavity mode couple to the Kittel mode, Figure \ref{APP_FIG_8} \textbf{a)} shows the magnitude of the $S_{21}$ parameter at resonance for this specific case and highlights the single anti-resonance between the two polaritonic branches. On the other hand, in the case where the YIG sphere is at a chiral position, only one of the two mode is coupled and the uncoupled cavity mode splits the anti-resonance in two (see Figure \ref{APP_FIG_8} \textbf{b)}).

\begin{figure}
    \centering
    \includegraphics[width=0.6\linewidth]{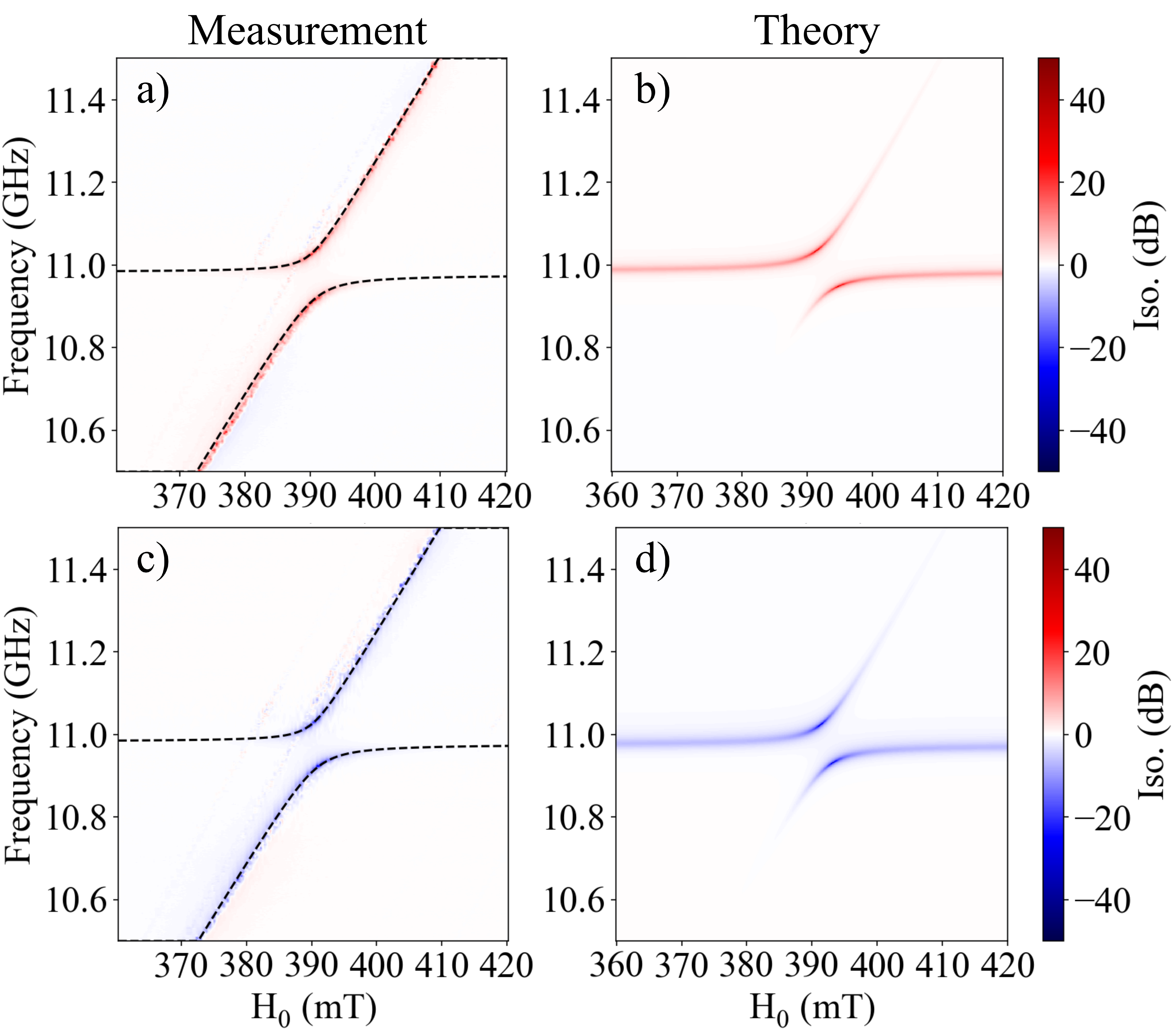}
    \caption{\textbf{a)}, \textbf{b)} and \textbf{c)}, \textbf{d)} Comparison between measurement and theory for a single YIG sphere at position $(\rho_+,0)$ and $(\rho_-,0)$ respectively. The black dashed lines correspond to the fitting using expression \ref{iso}.}
    \label{APP_FIG_9}
\end{figure}

Determining the two transmission parameters allows us to evaluate the isolation ratio following the definition:

\begin{equation}
    \text{Iso}.=20*\text{log}_{10}(S_{21}/S_{12}).
    \label{iso}
\end{equation}

from this expression we are able to fit the measurements and find good agreement between the experimental data and our theoretical model as it is showed in Figure \ref{APP_FIG_9}.

\subsection{Input-output model for two YIG spheres}\label{appendixB3}

Now that we have determined the expression of the transmission parameters for a single YIG sphere loaded in the cavity, we want to extend our theoretical model to the two YIG spheres case. For this situation, we must consider two bosonic operators: $\hat{m}_1$ and $\hat{m}_2$ instead of one, corresponding to the two magnetic samples inside the cavity. By taking into account this change, we are able to modify the free Hamiltonian such as:

\begin{equation}
    \frac{\hat{H}_{free}}{\hbar} = \Tilde{\omega}_+ \hat{\alpha}^{\dagger}_+ \hat{\alpha}_+ + \Tilde{\omega}_- \hat{\alpha}^{\dagger}_- \hat{\alpha}_- + \Tilde{\omega}_{K1} \hat{m}^{\dagger}_1 m_1 + \Tilde{\omega}_{K2} \hat{m}^{\dagger}_2 \hat{m}_2 + \text{H.c},
\end{equation}

with the complex eigenfrequencies $\Tilde{\omega}_{K1}=\omega_{K1}-i \kappa_{m1}/2$ and $\Tilde{\omega}_{K2}=\omega_{K2}-i \kappa_{m2}/2$ of the corresponding magnon modes. On the other hand the new interaction Hamiltonian for two YIG spheres is written:

\begin{equation}
    \frac{\hat{H}_{int}}{\hbar} = \prescript{1}{}{G}_+ \hat{\alpha}_+ \hat{m}^{\dagger}_1 + \prescript{1}{}{G}_- \hat{\alpha}_- \hat{m}^{\dagger}_1 + \prescript{2}{}{G}_+ \hat{\alpha}_+ \hat{m}^{\dagger}_2 + \prescript{2}{}{G}_- \hat{\alpha}_- \hat{m}^{\dagger}_2 + \text{H.c},
\end{equation}

where $\prescript{l}{}{G}_{\pm}$ is the complex coupling strength between the photonic mode either clockwise (+) or counterclockwise (-) and the $l$th magnon mode written as:

\begin{equation}
    \prescript{l}{}{G}_{\pm}=|g_{\pm}(\rho)|e^{\mp 3i\phi_l}e^{i\theta_\pm}
\end{equation}

with $\phi_l$ the azimuthal coordinate of the $l$th sphere in the polar basis and $\varphi_{\pm}$ the coupling phase. The $g_{\pm}$ terms are two function depending on the $\rho$ coordinate of the YIG sphere. The system of equations of motion after a Fourier transform for the double YIG spheres case is written:

\begin{align}
    \label{alpha-_2YIG}
    \omega \hat{\alpha}_- &= 
    \begin{aligned}[t]
        &\Tilde{\omega}_- \hat{\alpha}_- - \prescript{1}{}{G}^*_- \hat{m}_1 +  \prescript{2}{}{G}^*_- \hat{m}_2 \\
        &- i\kappa^*_{ccw,1}\left(b^{in}_1 + \tfrac{1}{2}\kappa_{cw,1} \hat{\alpha}_+ + \tfrac{1}{2}\kappa_{ccw,1}\hat{\alpha}_- \right) 
        - i\kappa^*_{ccw,2} \left(b^{in}_2 + \tfrac{1}{2}\kappa_{cw,2}\hat{\alpha}_+ + \tfrac{1}{2} \kappa_{ccw,2}\hat{\alpha}_- \right)
    \end{aligned} \\[1ex]
    \label{alpha+_2YIG}
    \omega \hat{\alpha}_+ &= 
    \begin{aligned}[t]
        &\Tilde{\omega}_+ \hat{\alpha}_+ - \prescript{1}{}{G}^*_+ \hat{m}_1 +  \prescript{2}{}{G}^*_+ \hat{m}_2 \\
        &- i\kappa^*_{cw,1}\left(b^{in}_1 + \tfrac{1}{2}\kappa_{cw,1} \hat{\alpha}_+ + \tfrac{1}{2}\kappa_{ccw,1}\hat{\alpha}_- \right)
        - i\kappa^*_{cw,2} \left(b^{in}_2 + \tfrac{1}{2}\kappa_{cw,2}\hat{\alpha}_+ + \tfrac{1}{2} \kappa_{ccw,2}\hat{\alpha}_- \right)
    \end{aligned} \\
    \omega \hat{m}_1 &= \Tilde{\omega}_{K1} \hat{m}_1 + \prescript{1}{}{G}_+ \hat{\alpha}_+ + \prescript{1}{}{G}_- \hat{\alpha}_- \\
    \omega \hat{m}_2 &= \Tilde{\omega}_{K2} \hat{m}_1 + \prescript{2}{}{G}_+ \hat{\alpha}_+ + \prescript{2}{}{G}_- \hat{\alpha}_-.
\end{align}

We isolate the the two magnon operators such that:

\begin{align}
    \label{m1_2YIG}
    \hat{m}_1= \frac{\prescript{1}{}{G}_+}{\omega-\Tilde{\omega}_{K1}}\hat{\alpha}_+ + \frac{\prescript{1}{}{G}_-}{\omega-\Tilde{\omega}_{K1}}\hat{\alpha}_- \\
    \label{m2_2YIG}
    \hat{m}_2= \frac{\prescript{2}{}{G}_+}{\omega-\Tilde{\omega}_{K2}}\hat{\alpha}_+ + \frac{\prescript{2}{}{G}_-}{\omega-\Tilde{\omega}_{K2}}\hat{\alpha}_-,
\end{align}

from this, we inject the two expressions \ref{m1_2YIG} and \ref{m2_2YIG} inside the expressions of \ref{alpha-_2YIG} and \ref{alpha+_2YIG}. In order to simplify the notations we write: $\Delta K1=\frac{1}{\omega-\Tilde{\omega}_{K1}}$ and $\Delta K2=\frac{1}{\omega-\Tilde{\omega}_{K2}}$. Finally, we obtain the following equation:

\begin{equation}
    (\omega-\Tilde{\omega}'_--(|\prescript{1}{}{G}_-|^2 \Delta K1 + |\prescript{2}{}{G}_-|^2 \Delta K2))\hat{\alpha}_- = ( \prescript{1}{}{G}^*_- \prescript{1}{}{G}_+ \Delta K1 +  \prescript{2}{}{G}^*_- \prescript{2}{}{G}_+ \Delta K2 -\frac{i}{2} J) \hat{\alpha}_+ - i(\kappa^*_{cw,1}b^{in}_1 + \kappa^*_{cw,2}b^{in}_2).
\end{equation}

\begin{figure}
    \centering
    \includegraphics[width=0.6\linewidth]{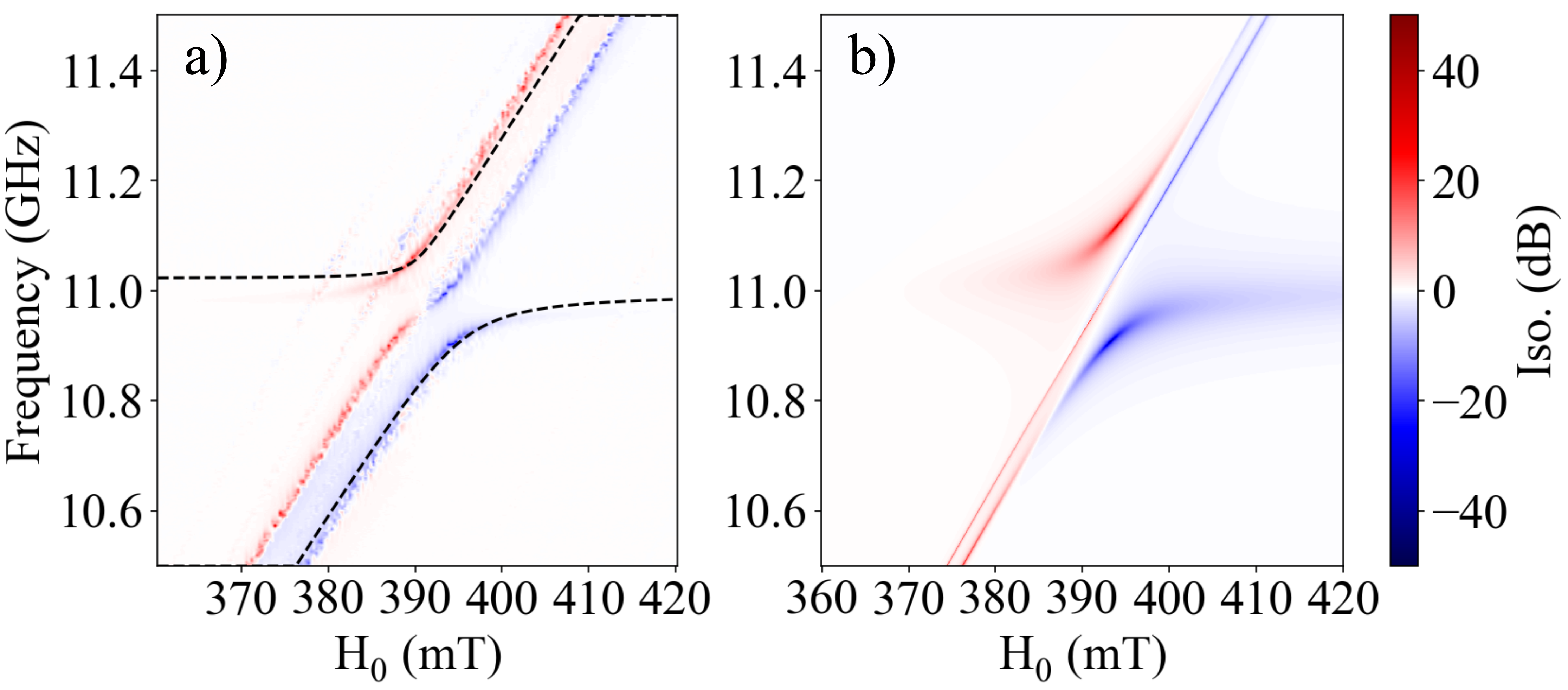}
    \caption{\textbf{a)} Measured and \textbf{b)} calculated isolation ratio for two YIG spheres loaded in the cavity at positions $(\rho_-, 0)$ and $(\rho_-, \pi)$. The black dashed lines are the fitting curves plotted by using \ref{S21_2YIG} and \ref{S12_2YIG}.}
    \label{APP_FIG_10}
\end{figure}

Then, by following the same steps as the single YIG case, we can extract the expression of the two transmission parameters for two YIG spheres loaded in the system:

\begin{equation}
    S_{21} = -i \tfrac{\kappa_{cw2}[\kappa_{ccw1}^*(\prescript{1}{}{G}^*_{+}\prescript{1}{}{G}_- \Delta K1 + \prescript{2}{}{G}^*_+\prescript{2}{}{G}_- \Delta K2-\frac{i}{2}J^*) + \kappa_{cw1}^*\Gamma'_-] + \kappa_{ccw2}[\kappa_{cw1}^*(\prescript{1}{}{G}^*_{-}\prescript{1}{}{G}_+ \Delta K1 + \prescript{2}{}{G}^*_{-}\prescript{2}{}{G}_+ \Delta K2-\frac{i}{2}J) + \kappa_{ccw1}^*\Gamma'_+]}{\Gamma'_-\Gamma'_+ - (\prescript{1}{}{G}^*_{-}\prescript{1}{}{G}_+ \Delta K1 + \prescript{2}{}{G}^*_{-}\prescript{2}{}{G}_+ \Delta K2-\frac{i}{2}J) (\prescript{1}{}{G}^*_{+}\prescript{1}{}{G}_- \Delta K1 + \prescript{2}{}{G}^*_+\prescript{2}{}{G}_- \Delta K2-\frac{i}{2}J^*)}
    \label{S21_2YIG}
\end{equation}

and :

\begin{equation}
    S_{12} = -i \tfrac{\kappa_{cw1}[\kappa_{ccw2}^*(\prescript{1}{}{G}^*_{+}\prescript{1}{}{G}_- \Delta K1 + \prescript{2}{}{G}^*_+\prescript{2}{}{G}_- \Delta K2-\frac{i}{2}J^*) + \kappa_{cw2}^*\Gamma'_-] + \kappa_{ccw1}[\kappa_{cw2}^*(\prescript{1}{}{G}^*_{-}\prescript{1}{}{G}_+ \Delta K1 + \prescript{2}{}{G}^*_{-}\prescript{2}{}{G}_+ \Delta K2-\frac{i}{2}J) + \kappa_{ccw2}^*\Gamma'_+]}{\Gamma'_-\Gamma'_+ - (\prescript{1}{}{G}^*_{-}\prescript{1}{}{G}_+ \Delta K1 + \prescript{2}{}{G}^*_{-}\prescript{2}{}{G}_+ \Delta K2-\frac{i}{2}J) (\prescript{1}{}{G}^*_{+}\prescript{1}{}{G}_- \Delta K1 + \prescript{2}{}{G}^*_+\prescript{2}{}{G}_- \Delta K2-\frac{i}{2}J^*)}
    \label{S12_2YIG}
\end{equation}

with $\Gamma'_{\pm} = \omega - \Tilde{\omega}'_{\pm} - (|\prescript{1}{}{G}_{\pm}|^2 \Delta K1 + |\prescript{2}{}{G}_{\pm}|^2 \Delta K2)$. Substituting the two expressions of the transmission parameters in the definition of the isolation ratio given by equation (\ref{iso}), we use it as a fitting function that matches the experimental data obtained for the cavity loaded with two YIG spheres at positions $(\rho_-, 0)$ and $(\rho_-, \pi)$, as it is shown in Figure \ref{APP_FIG_10}. It is important to note that a shift of the Kittel resonances for each magnon mode has been introduced to model the linear curves of nonzero isolation ratio in between the two branches. The latter confirms that the origin of those linear curves comes from the fact that the YIG spheres have slightly different resonance frequency from one another.

\section{Non-reciprocity measurements}\phantomsection\label{appendixC}

\begin{figure}[hbt!]
    \centering
    \includegraphics[width=0.9\linewidth]{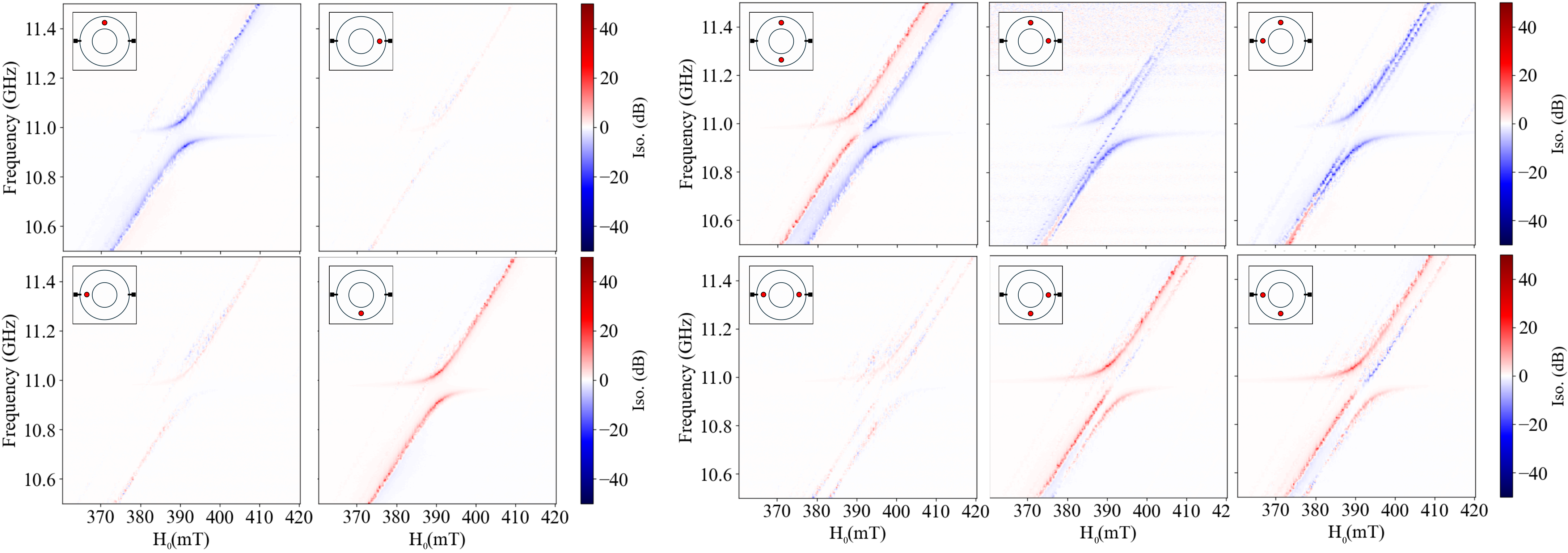}
    \caption{Isolation ratio according to frequency and bias magnetic field in dB for $N=1$ (left) and $N=2$ (right) YIG spheres loaded in the resonator at $\rho=\rho_-$. The insets show the configuration of the YIG spheres inside the resonator.}
    \label{APP_FIG_11}
\end{figure}

Figure \ref{APP_FIG_11} shows different position combination for one and two YIG spheres. For a single YIG sphere positioned at an angle of $\phi=\frac{\pi}{2}$ or $\phi=\frac{-\pi}{2}$. A rotation of $\frac{\pi}{2}$ will enable non-reciprocity favoring $S_{12}$ at position $(\rho_-,0)$ and $S_{21}$ at position $(\rho_-,\pi)$. This cancellation of the non-reciprocity is also observable when both YIG spheres occupy the two lateral positions as it is shown in the right panel of Figure \ref{APP_FIG_11}.

\begin{figure}[hbt!]
    \centering
    \includegraphics[width=0.7\linewidth]{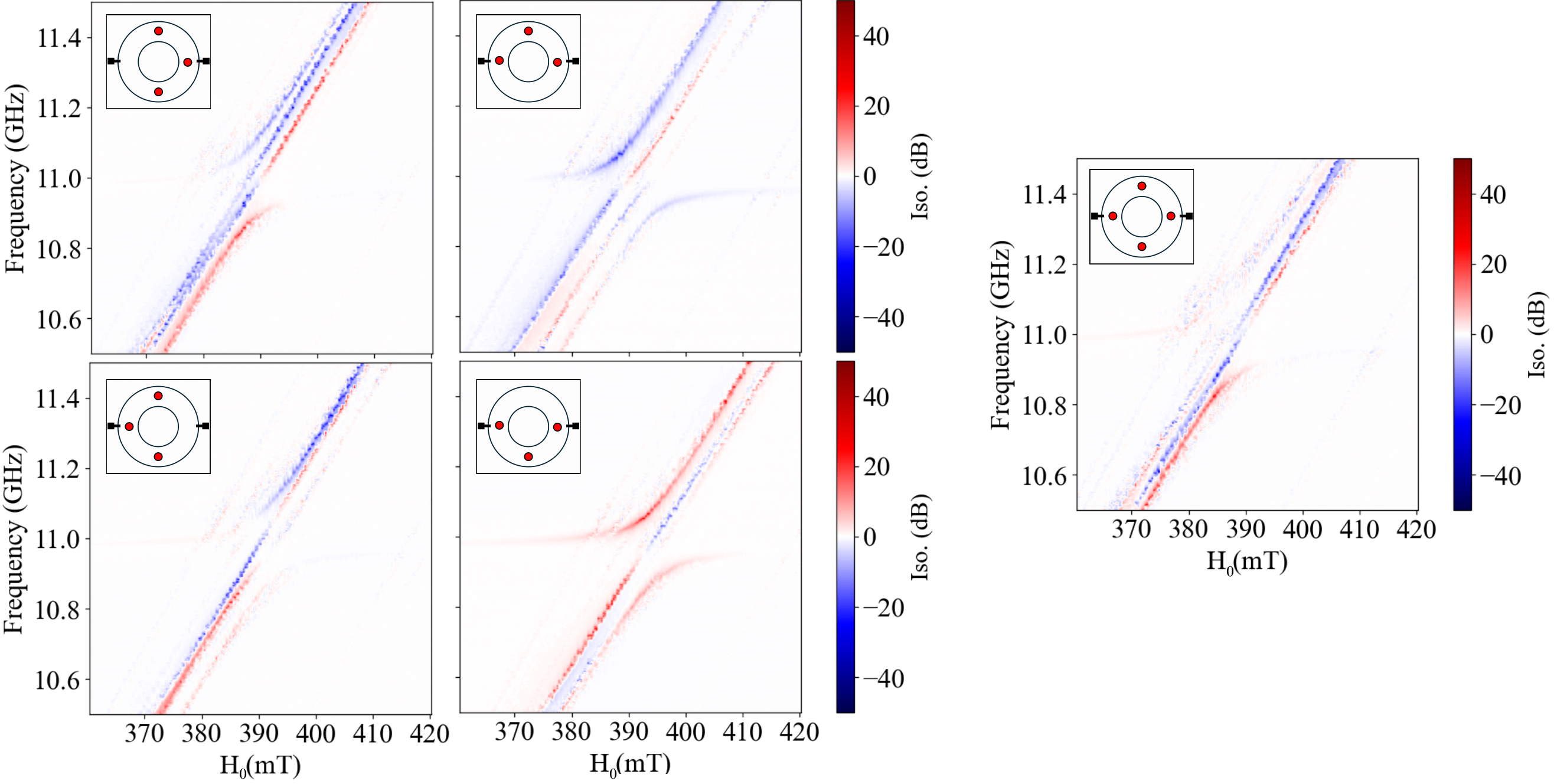}
    \caption{Isolation ratio for $N=3$ (left) and $N=4$ (right) YIG spheres loaded in the resonator at $\rho=\rho_-$.}
    \label{APP_FIG_12}
\end{figure}

Figure \ref{APP_FIG_12} displays the isolation ratio for $N=3$ and $N=4$. In those configurations we see how the properties of each individual YIG sphere is combining with each other. Each magnetic sample brings its contribution to the spectrum. For example, one can see that by adding a third YIG sphere to the lateral configuration showed in the right panel of Figure \ref{APP_FIG_11}, the isolation ratio increases, and the dominant transmission parameter is linked to the added YIG position ($\phi=0$ or $\phi=\pi$).

\end{document}